\documentstyle[aps,pre,twocolumn,epsf]{revtex}

\newcommand \scaleK {{\bar{K}}}                  
\newcommand \Curvapp {{\bar {\kappa}}}    
\newcommand \scs {{\bar{s}}}                 	 
\newcommand \us {{\bar{u}}}                      
\newcommand \Curv {{\bar{\kappa}}}               
\newcommand \MU {c \,}                           
\newcommand \Free {{\cal F}}
\newcommand \Vn {v}

\newcommand \free {{\cal F}\{\psi,c\}}

\newcommand \call[1] {{\cal L}_#1}

\newcommand \dqdt[1] {\frac{\partial #1}{\partial t}}
\newcommand \pqpq[2] {\frac{\partial #1}{\partial #2}}
\newcommand \pqps[2] {\partial #1/\partial #2}

\newcommand \fqfq[2] {\frac{\delta #1}{\delta #2}}

\newcommand \gpsi{\Gamma_{\psi}}   
 
\newcommand \gc {\Gamma_c}

\newcommand \goldp {\frac{\partial \psii}{\partial u}}

\newcommand \goldpus {\frac{\partial \psii}{\partial \us}}
\newcommand \goldcus {\frac{\partial \ci}{\partial \us}}

\newcommand \be {\begin{equation}}
\newcommand \ee {\end{equation}}
\newcommand \ben {\begin{eqnarray}}
\newcommand \een {\end{eqnarray}}
\newcommand \ep {\epsilon}

\newcommand \ffi[2] {f_i^{(#1,#2)}}
\newcommand \ff[2]  {f^{(#1,#2)}}
\newcommand \nline {\nonumber \\}
\newcommand \ci {c^{{\rm in}}_0}
\newcommand \co {c^{{\rm out}}_0}
\newcommand \cin[1] {\delta c^{{\rm in}}_#1}
\newcommand \con[1] {\delta c^{{\rm out}}_#1}
\newcommand \psii {\psi^{{\rm in}}_0}
\newcommand \psiib {\psi^{{\rm in}}}
\newcommand \psio {\psi^{{\rm out}}_0}
\newcommand \psiob {\psi^{{\rm out}}}
\newcommand \psiin[1] {\delta \psi^{{\rm in}}_#1}

\newcommand \mui {\mu^{{\rm in}}_0}
\newcommand \muo {\mu^{{\rm out}}_0}
\newcommand \muin[1] {\delta \mu^{{\rm in}}_#1}
\newcommand \muon[1] {\delta \mu^{{\rm out}}_#1}

\begin{document}
\draft

\twocolumn[\hsize\textwidth\columnwidth\hsize\csname 
@twocolumnfalse\endcsname

\title{Sharp interface limits of phase-field models}

\author{K. R. Elder$^1$, Martin Grant$^2$, Nikolas Provatas$^{2,3}$
and J. M. Kosterlitz$^{4}$}

\address{$^1$Department of Physics, Oakland University, Rochester, MI,
48309-4487}
\address{$^2$Physics Department, Rutherford Building, 3600 rue
University, McGill University,
Montr\'eal, Qu\'ebec, Canada H3A 2T8.}
\address{$^3$Pulp and Paper Research Institute of Canada, 570 St. Jean Blvd,
Montr\'eal, Qu\'ebec, Canada H9R-3J9.}
\address
{$^{4}$Department of Physics, Brown University, Providence,
RI 02912.}

\date{\today}

\maketitle

\begin{abstract}
The use of continuum phase-field models to describe the motion of 
well-defined interfaces is discussed for a class of 
phenomena, that includes order/disorder transitions, 
spinodal decomposition and Ostwald ripening, dendritic growth, 
and the solidification of 
eutectic alloys.  The projection operator method is used to 
extract the ``sharp interface limit'' from phase field models 
which have 
interfaces that are diffuse on a length scale $\xi$.  In particular,
phase-field equations are mapped onto sharp interface equations
in the 
limits $\xi \kappa \ll 1$ and $\xi v/D \ll 1$, where $\kappa$
and $v$ are respectively the interface curvature and velocity and
$D$ is the diffusion constant in the bulk.  The calculations 
provide one general set of sharp interface equations that 
incorporate the Gibbs-Thomson condition, the Allen-Cahn equation 
and the Kardar-Parisi-Zhang equation.

\end{abstract}


\vskip2pc]

\section{Introduction}

Many inhomogeneous systems involve domains of well-defined phases
separated by thin interfaces.  These include non-equilibrium systems
where phases are separating by spinodal decomposition or nucleation
and growth \cite{GUNTON}, or where solidification occurs by dendritic 
growth \cite{DENDRITES} or by the growth of eutectic crystals
\cite{EUTECTICS}. The phenomenological description of these phenomena
involves the motion of well-defined sharp interfaces.  The origin of 
such descriptions is often transparent, being obtained by symmetry
arguments and common sense.  Nevertheless the properties of 
sharp interface models can be quite subtle as is the case for
dendritic growth. 

	Unfortunately sharp interface models are difficult to 
simulate since this usually involves solving a diffusion equation 
subject to {\it moving} boundary conditions at the interfaces.  
A more convenient approach is to simulate models which 
describe the bulk phases as well as the interface structure.  While 
these models are wasteful in terms of simulating bulk regions\cite{pgd98}, 
no explicit boundary tracking is needed. This is the key element to a 
popular method for studying systems out of equilibrium, called 
``phase-field'' modeling.  In such an approach one or more
continuous fields which are functions of space $\vec r$ and time
$t$, are introduced to describe the phases present.  Typically 
these fields vary slowly in bulk regions and rapidly, on
length scales of the order of the correlation length, $\xi$,
near interfaces.  The free energy functional ${\cal F}$ 
determines the phase behavior and, with the equations of motion, gives 
a complete description.  In other contexts, such as critical 
dynamics \cite{GUNTON,HOHENBERG}, the fields are the order parameters
distinguishing the phases.  In a binary alloy, for example, the local
concentration or sublattice concentration can be described by 
such fields.  The ideas involved in this approach have a long history, 
going back to van der Waals \cite{VANDERWAALS}.  
Within the materials community, the use of continuum field
models is associated particularly with the work of Cahn and
collaborators \cite{CAHN,ALLENCAHN}.

Phase field models provide a complete microscopic description,
but are not necessarily appropriate for a particular system.
These models apply to a large number of microscopic systems
only when $\xi$ is much larger than any particular microscopic 
length, such as the lattice spacing, involved in the surface 
structure. Interpreted in the sense of describing 
the universal features of many microscopic models, $\xi$ is a 
mesoscopic length representative of microscopic structure.  
In a similar manner sharp-interface models apply to a large 
number of continuum field models of pattern formation or 
phase separation in the limit that the length 
scales defined by the patterns are much larger than 
$\xi$.  An important difference exists however in the 
construction of the two approaches.  Standard sharp-interface 
models are constructed from phenomenological descriptions 
of interfaces, while phase field models can be constructed 
to explicitly obey the fundamental principles of statistical 
mechanics.

	While continuum phase field models provide a 
fundamental approach which is clear and workable, it is important 
to establish the connection between this description and the sharp 
interface description.  The main difficulty which arises is 
how to take account of the finite thickness $\xi$ of the 
diffuse interface of the continuum model.  There has been 
a great deal of discussion in the literature on how this process 
is to be undertaken.  Some workers have extracted the
interface equations by taking the limit where the interface 
width of the phase field model goes to zero \cite{caji} for 
the Stefan problem of a pure material.  This approach is 
not very useful since the interface width is always finite.
More recently these calculations have been extended for 
special choices of the free energy functional ${\cal F}$ to 
include an interface of non-zero width\cite{karm,Alm}.

The purpose of this paper is to provide a clear and 
detailed derivation of the sharp interface equations. 
The approach follows the 
projection operator method of Kawasaki 
and Ohta \cite{KYOZI}, is generally applicable and 
eliminates the counter intuitive necessity 
of limiting the derivation of the sharp interface limit 
to some artificial form of the free energy 
functional ${\cal F}$ \cite{karm}.  This general 
calculation provides one set of equations that 
relate the parameters of the phase field equations 
to those of the sharp interface equations for a 
broad class of phenomena including
order-disorder transitions, dendritic growth, 
phase separation in binary alloys, eutectic growth, 
and surface roughening.  In particular, thermodynamic 
consistency is automatic in the present approach for 
nonzero interface widths.  This is in contrast to 
the unphysical approach of taking the limit 
of a zero interface width, which requires fine tuning of 
the free energy to obtain a thermodynamically consistent 
theory.

	The structure of the paper is as follows. 
In the next section a general introduction to phase 
field and sharp interface equations is given.  This 
is followed by a detailed calculation of the 
sharp interface equations from a set of phase field equations.
The usefulness of the sharp interface equations is then 
illustrated by considering the linear stability of planar 
interface separating various phases. 

\section{Describing Interface Phenomena using 
Phase-Field Models}

To see the relationship between the two approaches, it is useful to
construct the simplest equilibrium and non-equilibrium descriptions
of surfaces by both methods.
The main ideas developed below appear in the more general
case discussed in the following sections.
These ideas are present in work by
Allen and Cahn 
\cite{ALLENCAHN}
on the motion of anti-phase boundaries in the
kinetics of an order-disorder transition in a binary alloy.
This work recovers
the sharp-interface description from the phase-field model
in the appropriate limit {\em and\/}
predicts an experimentally testable consequence of a 
finite $\xi$.

First, consider a surface in equilibrium.  The
reason why droplets are spheres and interfaces are locally
flat is that these shapes minimize surface area.  The excess free energy
of a surface is proportional to its area $A$
\begin{equation}
\label{EQ:SURF}
\Delta {\cal F} =\sigma A 
\end{equation}
where the
proportionality constant $\sigma$ is the surface tension.
This simple common sense approach is the sharp interface model.

Contrast this with the phase field approach.  The free energy
functional ${\cal F}(\psi)$ for the scalar order parameter $\psi(\vec r)$
has the following familiar form, consistent with reflection symmetry ${\cal F}
(\psi) = {\cal F}(-\psi)$
\begin{equation}
\label{EQ:FREE}
 {\cal F}\{ \psi \} = 
 \int d\vec r \Bigl[\frac{1}{2}K_{\psi} (\vec\nabla \psi)^2 + f(\psi)
\Bigr], \end{equation}
where $K_{\psi}>0$ so that the square gradient interaction gives the free 
energy cost for inhomogeneities and the local bulk free energy has a 
double well form satisfying $f(\psi) = f(-\psi)$. 
An example is the $\psi^4$ form
\begin{equation}
\label{EQ:BULKFREE}
f = -\frac{a}{2}\psi^2 + \frac{b}{4}\psi^4,
\end{equation}
where  $a \propto (T_o - T) >0$ with $T_o$ the mean field
critical temperature and $b>0$ is a constant.

The motivation for employing such free energy functionals 
is as follows.  First, the free energy functional 
is constructed of local or bulk terms, $f(\psi)$, 
which interact through the gradient term.  Secondly  
${\cal F}$ must be an analytic function since it 
describes a microscopic system and should not 
be confused with the true thermodynamic free energy, 
$F \equiv -k_BT\,\ln \sum_{\psi} \exp(-{\cal F}/k_B T)$ 
where $k_B$ is Boltzmann's constant, 
which is non analytic at a phase transition\cite{GOLDENFELD}.  
The square gradient term is the simplest way for
the model to be well defined on small length scales.  Higher-order
gradient terms, such as $K_4 (\nabla^2 \psi)^2$ could, in principle,
describe correlations on still smaller scales, such as the 
internal structure of
the interface.  This is analogous to what is described by the
Percus-Yevick or hypernetted-chain theories of dense fluids
\cite{LIQUIDS}.  In the
absence of any specific experimental motivation, 
however, terms proportional to $K_{4}$ may be ignored.
From the point of view of
identifying and studying a well defined microscopic model, the
square gradient free energy provides a complete description.
Alternatively, the square gradient theory describes a large class of
microscopic models provided we consider mesoscopic scales 
$\xi(K)/a_o \gg 1$, where $a_o$ is a lattice constant.

It is particularly convenient that the microscopic phase field
description is so ``close'' to mean field theory.  In that
approximation, $f$ is simply the bulk free energy density.
In practice, the form of the free energy functional ${\cal F}$
can be constructed straightforwardly with reference to the phase
diagram of interest.  While one can also construct microscopic
 {\em lattice gas\/} models of phenomena such as phase separation and
dendritic solidification  \cite{DIK},
using similar arguments of universality and simplicity,
such models do not have this convenient feature.

If fluctuations are small, the equilibrium behavior of the
model is determined by the mean field approximation
\begin{equation}
\label{EQ:MEANFIELD}
\frac{\delta {\cal F}}{\delta \psi} =  -K_{\psi}\nabla^2 \psi +
\pqpq{f}{\psi} = 0.
\end{equation}
The homogeneous bulk solutions, valid well below $T_c$, are
given by $\partial f/\partial \psi_{eq} = 0$ and are equal to,
\begin{equation}
\label{EQ:PSIEQ}
\psi = \pm \psi_{eq} = \pm\sqrt{a/b}.
\end{equation}
for the bulk free energy given in Eq. (\ref{EQ:BULKFREE}).
Fluctuations around the bulk solutions satisfy
\begin{equation}
\label{EQ:CORREQ}
\langle (\psi(\vec r) -\psi_{eq})(\psi(0)-\psi_{eq}) \rangle
\sim e^{-r/\xi}
\end{equation}
for $r\gg\xi$, where $\xi = \sqrt{K_{\psi}/2a}$ and 
$\langle \cdots \rangle$ represents an ensemble average.

Now consider a system with a flat interface located at 
$y=0$ when $\psi(\vec r)$ depends only on $y$ and the interface profile
$\psiib(y)$ is the solution of
\begin{equation}
\label{EQ:PSIY}
-K_{\psi}\pqpq{^2\psiib}{y^2} + \pqpq{f}{\psiib} =0.
\end{equation} 
with $\psiib(\pm\infty) = \pm\psi_{eq}$.
Solving by quadratures gives
\begin{equation}
\label{EQ:CAHNTHINGY}
\frac{1}{2} K_{\psi}\Bigr(\pqpq{\psiib}{y} \Bigl)^2 = f(\psiib)
\end{equation} 
from which Eq. (\ref{EQ:SURF}) follows immediately. In contrast to the sharp
interface approach, this yields an explicit form for the surface tension
\begin{equation}
\label{EQ:SIGMA}
\sigma_{\psi} = K_{\psi}\int dy \left(\pqpq{\psiib}{y}\right)^2.
\end{equation}
For the particular form of ${\cal F}(\psi)$ of Eq. (\ref{EQ:FREE}) and
Eq. (\ref{EQ:BULKFREE}), one finds
\begin{equation}
\label{EQ:TANH}
\psiib (y) = \psi_{eq} \tanh\left(\frac{y}{2\xi}\right)
\end{equation}
and
\begin{equation}
\label{EQ:SIGMAPSI4}
\sigma_\psi = \frac{2K_{\psi}\psi_{eq}^2}{3\xi}
\end{equation}
The contrast between the sharp-interface model and the
phase-field model is revealing.  Both give the same macroscopic
description but the phase field approach gives a complete description
down to the smallest length scale.     

Now consider a sharp interface that is in local but not global 
equilibrium due to a gentle curvature. 
For simplicity consider phenomena where the order parameter is
not conserved, such as occurs when a binary alloy undergoes an 
order-disorder transition and $\psi$ is the sublattice concentration.
The interface moves to locally reduce the surface area and surface free
energy with an interface velocity $\Vn$.  Expanding $\Vn$ in a Taylor
series in powers of $\kappa$ gives
\begin{equation}
\label{EQ:VK}
\Vn = -\nu \kappa
\end{equation}  
to lowest order in the curvature.  This is the sharp interface
theory for the motion of anti-phase boundaries.  Note that, since $\nu$ is
the only coefficient which enters the theory and has dimensions of a
diffusion constant, any time-dependent length $R(t)$ must satisfy
\begin{equation}
\label{EQ:GROWTH}
R(t) = (\nu t)^{n},
\end{equation}
by dimensional analysis, where the growth exponent
\begin{equation}
\label{EQ:LAW}
n = {1}/{2}.
\end{equation}
Such an approach was first done by Lifshitz \cite{LIFSHITZ}
and by Turnbull \cite{TURNBULL}.
The sharp interface treatment alone cannot predict the
value of $\nu$.  An additional argument, which turns out to be incorrect
for the motion of antiphase boundaries, was used to 
predict $\nu \propto \sigma_\psi \Gamma_{\psi}$,
where $\Gamma_{\psi}$ is a mobility.

A first principles approach to this phenomenon is due to Allen and Cahn
\cite{ALLENCAHN}.
Neglecting noise, the equation of motion for the non-conserved
sublattice concentration is \cite{GUNTON,HOHENBERG,ALLENCAHN}
\begin{equation}
\label{eq:start}
\frac{\partial \psi}{\partial t} = 
       - \Gamma_{\psi} \frac{\delta {\cal F}}{\delta \psi} =
- \Gamma_{\psi} \Bigl[-K_{\psi}\nabla^2 \psi +\frac{\partial f}{\partial
\psi} \Bigr].
\end{equation}
Allen and Cahn denoted the position of the antiphase boundary
by a curved, time-dependent
interface $u(\vec r, t) = 0$.  They then looked for solutions of the form 
$\psi (\vec r , t) = \psiib (u(\vec r, t))$.   This gives
\begin{equation}
\label{EQ:CAHNTRICK}
\frac{\partial \psi}{\partial t} = -\Vn \pqpq{\psiib}{u}
= - \Gamma_{\psi} \Bigl[
-K_{\psi}\pqpq{^2 \psiib}{u^2} 
-K_{\psi}\kappa \pqpq{\psiib}{u} +\pqpq{f}{\psi} \Bigr],
\end{equation}
where $\kappa = -\vec\nabla \cdot \hat n$ with $\hat n = \vec
\nabla u/|\vec\nabla u|$ the unit vector normal to the interface and $u(\vec
r,t)$ a coordinate in the direction $\hat n$.
Eliminating $\partial f/\partial \psiib$ using Eq.\ (\ref{EQ:PSIY}) gives 
Eq. (\ref{EQ:VK}) where, in contrast to the sharp interface 
theory, one obtains an explicit expression for the transport 
coefficient,
\begin{equation}
\label{EQ:MOBILITY}
  \nu = K_{\psi}\Gamma_{\psi}.
\end{equation}
Results in the presence of stochastic noise
have been obtained by many authors, particularly
Bausch, Domb, Janssen and Zia
\cite{ROYCE}, 
and by Kawasaki and Ohta 
\cite{KYOZI}.

Although both
approaches correctly find that the normal velocity is proportional to
the curvature, the Allen-Cahn result for $\nu$ \cite{ALLENCAHN} is noteworthy.
The earlier theory, which argued $\nu \propto\sigma_\psi \Gamma_{\psi}$
implies a strong dependence of the velocity $\Vn$ on interfacial thickness
since $\sigma\propto 1/\xi$, from Eq.\
(\ref{EQ:SIGMAPSI4}).  
In contrast, Allen and Cahn predict that $\nu$ is independent of
interfacial thickness.  This was clearly
demonstrated in an experiment by Pindak {\it et al.\/}\
\cite{PINDAK} where they studied orientational patterns in
freely suspended dipolar smectic C liquid crystal films.
Since their smectic C films have a permanent electric
dipole moment of magnitude $P$, the director angle $\phi$ can be
oriented with an electric field of magnitude $E$.  The
free energy is 
\begin{equation}
\label{EQ:LIQCRYS}
 {\cal F} = 
\int d\vec r \Bigl[\frac{1}{2} K(\nabla \phi)^2 -EP \cos \phi\Bigr],
\end{equation}
so the width of the interface, 
\begin{equation}
\label{EQ:XIE}
\xi \propto 1/\sqrt E,
\end{equation}
can be varied easily.
The experiments directly verify that the size of a domain of stable
orientation grows as $R(t) = (\nu t)^{1/2}$,
where $\nu$ is independent of interface width in accord with the
prediction of Allen and Cahn \cite{ALLENCAHN}.  In addition, the experiments
show that the
sharp interface result, $v \propto \kappa$, is independent of
interface width provided $\xi/R \ll 1$.

It is straightforward to include an external field $\MU$ coupling
linearly to $\psi$.  Then ${\cal F}\rightarrow{\cal F} +
{\cal F}_{ext}$ with
\begin{equation}
\label{EQ:EXTFIELD}
\Free_{ext} =  - \int d\vec r \MU  \psi
\end{equation}
and one of the phases becomes metastable depending on the sign of $c$,
so that the interface translates even if it is flat. Hence
\begin{equation}
\label{EQ:KPZ1}
\Vn = v_{0}(c) - \nu \kappa
\end{equation}
with $v_{0}(c)\propto c$.
This allows one to simulate Kolmogorov-Avrami-Johnson-Mehl growth
of droplets \cite{AVRAMI} and, in the presence of noise,
Kardar-Parisi-Zhang dynamic roughening \cite{KPZ}.

Other processes can be simulated when one of the phases is
metastable and the growth of the stable from the metastable phase is
controlled by a conservation law.
This describes solidification of a metastable supercooled liquid phase
and the growth of the stable solid is limited by the diffusion of latent
heat from the surface of the moving solid front.
The external field $\MU$ is then proportional to the latent heat. In the
sharp interface formulation, $\MU$ obeys a diffusion equation in the
bulk phases
\be
\label{EQ:STEADY}
\frac{\partial c}{\partial t} = D_{c} \nabla^2 \MU,
\ee
where $D_{c}$ is a diffusion constant. 

The steady state velocity $v$ of the interface is given by integrating
Eq. (\ref{EQ:STEADY}) across the interface to obtain
\be
\label{EQ:BC-Vn}
\Vn \propto\hat n\cdot(D_c^{+}\vec\nabla c|^{+} -D_c^{-}\vec\nabla c|^{-})
\ee
where the superscripts $\pm$ refer to the values of the
normal gradient of $\MU$ on either side of the interface.
The condition of local equilibrium at the interface is 
\be
\label{EQ:LOCALEQ}
\delta\MU\propto\kappa
\ee
which is a Gibbs-Thompson condition relating the local excess concentration
$\delta c$ to the curvature. This says that
excess external driving force is balanced by the curvature $\kappa$.
In regimes of high undercooling, this is
sometimes supplemented by an additive term proportional to $\Vn$,
describing kinetic undercooling.

To study this by a phase-field approach,
let $\MU $ be a continuous function of space and
time which is conserved:
\begin{equation}
\label{EQ:CONSEXTFIELD}
\frac{\partial \MU(\vec r, t)}{\partial t} = \Gamma_{c} \nabla^2
\frac{\delta \Free}{\delta\MU}
\end{equation}
where $\Gamma_{c}$ is the mobility of the field $c$. For the 
model to be well defined, a self energy for $\MU$ or 
a positive additive contribution to $\Free$ of the form
\begin{equation}
\label{EQ:SELFENERGY}
\Free_{\MU}  \propto \int d\vec r \, \left[\frac{1}{2} \MU^2 \right]
\end{equation}
must be included.
Within a mean-field approximation, this gives a homogeneous
equilibrium solution $\MU \propto \psi$.  Note that the
interface invades the metastable phase because of $\Free_{ext}$
but it must also satisfy the conservation of $\MU$ as defined by
Eqs. (\ref{EQ:CONSEXTFIELD}) and (\ref{EQ:SELFENERGY}). This implies 
that the interface deforms into a parabolic shape, dumping
excess $\MU$ to the sides while propagating forward at a constant
velocity.  The parabolic shape has a constant growth velocity in
the forward direction, satisfying the fact the system is driven with a
constant thermodynamic force, while lateral growth has a velocity $\sim
t^{-1/2}$, thereby satisfying the conserved diffusion equation  for
$\MU$.

It turns out that, although this is the right approach,
the implementation needs
some fine tuning.  First, when considering dendritic growth, the theory
of microscopic solvability \cite{MICROSOLV} has shown that dendrites require
an anisotropic surface tension to be well defined.  Hence, one must let
$K \rightarrow K(\vec \nabla \psi/|\nabla \psi|)$ in some convenient 
prescribed way.
Next, in a very useful paper,
Kobayashi \cite{KOBAYASHI} has noted that, to keep the
equilibrium solutions $\psi_{eq}$ from shifting appreciably when
$\MU$ is applied, ${\cal F}_{ext}$  of Eq. (\ref{EQ:EXTFIELD}) should be
modified to
\begin{equation}
\label{EQ:KOBAYASHI}
\Free_{ext} = - \int d \vec r \MU \Psi(\psi)
\end{equation}
where $\Psi(\psi)$ is an odd function of $\psi$ satisfying
\begin{equation}
\label{EQ:KOBAYASHI2}
\partial \Psi/\partial \psi_{eq} = 0. 
\end{equation}
For example, if $\psi_{eq} = \pm 1$, one can choose 
$\partial \Psi /\partial \psi  = (1 - \psi^2)^{2}$.  Other forms are possible.

Finally, one can choose to make the $\MU$ field phase separate by
replacing the self-energy term in the free energy of
Eq.\ (\ref{EQ:SELFENERGY}) with a double well form, analogous to
Eq.\ (\ref{EQ:BULKFREE}).  This permits the study of eutectic
crystallization \cite{EUTECTICS}.  
In the remainder of this paper a detailed implementation of these 
ideas is given, making connections to the sharp interface limit. 

\section{Equilibrium}
\label{sec:equilibrium}

	Consider two fields,
a non-conserved phase field $\psi$ and a conserved field $c$.  
The phase field distinguishes between, for example, liquid and
solid phases and the $c$ field can be taken as a concentration.
The free energy functional describing the system can be written as
\be
\label{EQ:FREEEN}
\free = \int d\vec{r} \left[ \frac{1}{2}K_{\psi}|\vec{\nabla}\psi|^2 +
\frac{1}{2}K_c|\vec{\nabla}c|^2 +f(\psi,c) \right]
\ee 
where $f(\psi,c)$ is the local bulk free energy density and 
the gradient terms account for interfaces and inhomogeneities as discussed
above. The dynamics of these fields are described by the equations of motion
for the non-conserved $\psi$,
Eq.\ (\ref{eq:start})
and the conserved concentration $c$,
\ben
\label{eq:start2}
\dqdt{c} &=& \gc \nabla^2 \fqfq{\Free}{c}
=\gc \nabla^2 
\left[- K_c \nabla^2 c + \pqpq{f}{c}\right].
\een
where $\delta \free /\delta c = \mu$ is a chemical potential and
$\gc$ is a mobility.
The usual additive noise terms, related to the
transport coefficients $\Gamma_{c,\psi}$ by fluctuation-dissipation
relations \cite{GUNTON,HOHENBERG}, have been neglected for
simplicity. In mean field theory, the equilibrium states of the system
with an interface normal to $\hat y$ are
defined by 
\be
\label{eq:eqpsi}
K_{\psi}\pqpq{^2\psi}{y^{2}} = \pqpq{f}{\psi} 
\ee
and 
\be
\label{eq:eqc}
K_{c}\pqpq{^2c}{y^{2}} = \pqpq{f}{c}-\mu_{eq} 
\ee
where $\mu_{eq}$ is the chemical potential of the uniform equilibrium
states. Integrating Eq. (\ref{eq:eqc}) over $c$ gives the Maxwell's equal
area construction rule, 
$\int_{c_1}^{c_2} dc \ \left(\partial f/\partial c - \mu_{eq}\right) = 0$,
where $c_1$ and $c_2$ are defined
by $\partial f/\partial  c|_{c_1,c_2} = \mu_{eq}$.  

\begin{figure}[btp]
\begin{minipage}{8.0cm}
\epsfxsize=8.0cm \epsfysize=8.0cm
\epsfbox{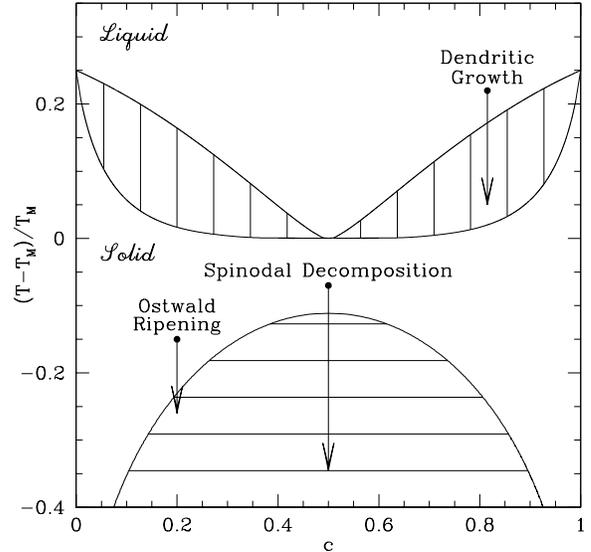}
\end{minipage}
\caption{Mean field phase diagram obtained from the bulk free energy of 
Eq. (\ref{eq:bulk}) for the parameters $\alpha=\beta=1.0$ and 
$u=0.6$.  
In this figure the regions containing vertical and horizontal 
lines are liquid/solid and solid/solid coexistence regions respectively.}
\label{fig:liqsol}
\end{figure}

\begin{figure}[btp]
\begin{minipage}{8.0cm}
\epsfxsize=8.0cm \epsfysize=8.0cm
\epsfbox{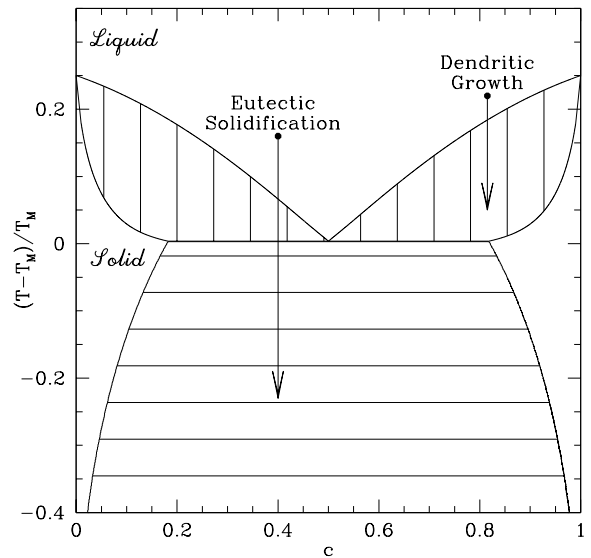}
\end{minipage}
\caption{Mean field phase diagram obtained from bulk free energy given 
in Eq. (\ref{eq:bulk}) for the parameters $\alpha=\beta=1.0$ and 
$u=0.45$.
In this figure the regions containing vertical and horizontal 
lines are liquid/solid and solid/solid coexistence regions respectively. }
\label{fig:eut}
\end{figure}

For a generic $f(c,\psi)$ there may be many possible equilibrium and
metastable states contained in this free energy.  For illustration, 
consider the following bulk free energy
\ben
\label{eq:bulk}
f(\psi,c) &=&
 u \frac{T}{T_M} \left[c \ln c + \left(1-c\right) \ln \left(1-c\right) \right]
\nonumber \\
&+&\left(\alpha\Delta T - \beta(c-\frac{1}{2})^{2}\right)\Psi(\psi) - \frac{1}{2}\psi^{2} + \frac{1}{4}\psi^{4}.
\een
where $\Psi(\psi)\equiv 2\psi - 4\psi^{3}/3 + 2\psi^{5}/5$, $\Delta T \equiv (T-T_M)/T_M$ with $T_M$ the melting temperature
and the other phenomenological parameters are determined by matching to
experimental phase diagrams.  
If these parameters are chosen as
$\alpha=\beta=1.0$ and $u=0.6$, the mean-field phase diagram shown in Fig.
(\ref{fig:liqsol}) emerges.  As can be seen, this phase diagram contains
liquid/solid and solid/solid coexistence regimes.
For this symmetric free energy the melting temperature at $c=1/2$ 
is denoted $T_M$ and the
critical point of the solid/solid coexistence regime is at $(c,T) =
(1/2,T_{c})$ with $T_{c}<T_{M}$. 
As the parameter $u$ is decreased, $T_{c}$ increases until the solid/solid
coexistence region collides with the liquid/solid coexistence regime
when $T_{c} > T_{M}$ and a eutectic point is formed at $(c,T) =
(1/2,T_{E})$ as shown in Fig. (\ref{fig:eut}).

	As can be seen from the phase diagrams of
Figs. (\ref{fig:liqsol}) and (\ref{fig:eut}), this simple free energy 
contains many phases and, in conjunction with appropriate equations
of motion, can be used to study a wide variety of phenomena.  A number
of `quenches' have been highlighted on these diagrams to illustrate
several different kinetic processes that may arise. 

\noindent 
A quench is defined 
as a rapid change in temperature which takes a system from one region 
of the phase diagram to another and is often considered instantaneous
in theoretical modeling.  In the next section
the dynamics of an interface separating a stable and a metastable 
phase is considered.  The calculations are done in a general
manner to include all the possible quenches shown in Figs.
(\ref{fig:liqsol}) and (\ref{fig:eut}).
To construct the relevant sharp interface
equations no reference will be made to the explicit form of the
bulk free energy term $f$.  It is assumed that
$f$ has been chosen merely so that all the phases of interest are 
well defined.

\section{Non-equilibrium}

The goal of this section is to derive the sharp 
interface equations for a systems described by a free energy 
functional ${\cal F}$ (such as is given by 
Eqs. (\ref{EQ:FREEEN}) and (\ref{eq:bulk})) and 
the Langevin equations given 
in Eqs. (\ref{eq:start}) and (\ref{eq:start2}). 
These latter
equations should be supplemented by additive stochastic noises,
$\eta_\psi(\vec{r},t)$ and 
$\eta_c(\vec{r},t)$ of zero mean and correlations 
$<\eta_{\psi}(\vec r,t)\eta_{\psi}(\vec r\,',t')> =
2\Gamma_{\psi}T\delta(\vec r-\vec r\,')\delta(t-t') 
$
and 
$
<\eta_{c}(\vec
r,t)\eta_{c}(\vec r\,',t')> = 2\Gamma_{c}T\nabla^{2}\delta(\vec r-\vec
r\,')\delta(t-t') 
$
with $<\eta_{\psi}\eta_{c}> =0$ as required by the
fluctuation dissipation theorem. 
With these stochastic noises, the dynamical
equations Eq. (\ref{eq:start}) and Eq. (\ref{eq:start2}) are the simplest
equations which respect the macroscopic conservation laws and also ensure that 
the system evolves towards its ultimate state of thermal and mechanical
equilibrium with its external environment. 

The asymptotic analysis proceeds by expanding around a
planar equilibrium interface described by Eqs. (\ref{eq:eqpsi}) and
(\ref{eq:eqc}).  The interface can be taken out of equilibrium by 
either gently curving it or by making one of the two bulk
phases metastable.  In the former case, a gentle curvature 
is one in which the radius of curvature $1/\kappa$ is large compared to the
interface width or correlation length.  Thus one small {\it
dimensionless} expansion parameter is $\kappa \xi$.  In the 
latter case, the difference between the
free energy of the stable and metastable phases causes the interface to 
propagate into the metastable phase.  If the free energy difference 
is small the propagation velocity $v$ is small. In this context,
a small velocity means that the interface moves so slowly that 
a steady state diffusion field is allowed to form in front of the
interface.  In other words, the time for the diffusion field 
to relax when the interface moves a distance $\xi$
should be much smaller than the time $\xi/v$ taken for the interface
to move that distance.  Since the diffusion time $\tau = \xi^{2}/D$,
this leads to another small {\it dimensionless} parameter $\xi v/D$ 
which is known as the interface P\'{e}clet number.  In the following analysis the
interface equations will be obtained to lowest order in both 
small parameters.  Technically the expansion to lowest order in both 
small parameters can be achieved if they are regarded as
the same order in the expansion.  In the calculations to follow 
both parameters will be taken to be ${\cal O}(\epsilon)$ with $\epsilon
<<1$.

\begin{figure}[btp]
\begin{minipage}{8.0cm}
\epsfxsize=8.0cm \epsfysize=8.0cm
\epsfbox{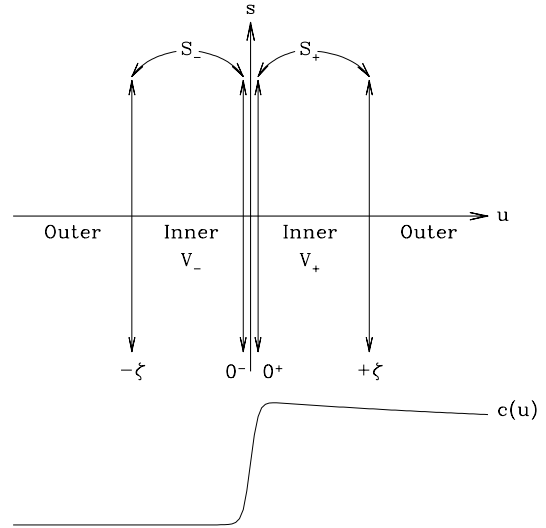}
\end{minipage}
\caption{Illustration of inner and outer regions used in 
computations.}
\label{fig:regions}
\end{figure}

	The calculations make use of the fact that the fields behave
qualitatively different close and far from the interface.
In the region close to the interface, the fields  vary
rapidly over distances ${\cal O}(\xi)$ while, far from the interface,
they vary over distances ${\cal O}(\xi/\epsilon)$.  
If there exists a length scale $\zeta$ such that
$1 << \zeta/\xi << 1/\epsilon$, then distinct `inner' and
`outer' regions can be defined, as shown in Fig. (\ref{fig:regions}) and it 
is appropriate to solve in both inner and outer regions 
and match the solutions at the length scale $\zeta$.
Formally, the technique requires an inner expansion
near the interface and an outer expansion far from
the interface.

\subsection{Inner Expansion}
\label{sec:innerexp}

	Consider an inner region defined by $-\zeta < u <+\zeta$, where
$u$ is a coordinate normal to the interface and $1 << \zeta/\xi <<
1/\epsilon$. The aim is to obtain asymptotic expansions for the solutions to
the evolution equations (\ref{eq:start}) and (\ref{eq:start2}) valid in this
inner region. The latter can be written in a compact form as
\be
\label{eq:start2a}
\frac{1}{\Gamma_{c}}\frac{\partial c}{\partial t} = \nabla^{2}\delta\mu
\ee
where
\be
\label{eq:deltamu}
\delta\mu = \mu(\vec r) - \mu_{eq} = -K_{c}\nabla^{2}c + \frac{\partial
f}{\partial c} - \mu_{eq}
\ee

The first step is to partition the system into two regions $V_{+}$ and
$V_{-}$ bounded by surfaces $S_{+}$ and $S_{-}$ respectively.
The region $V_{+}$ is defined by $0 < u < \zeta$ and  
and similarly for $V_{-}$.  The position of the interface between 
two bulk regions is defined as $u(\vec r,t) = 0$. 
These definitions are purely formal,
but to fix ideas, the surface $u(\vec r,t) = 0$ may be 
regarded as the surface near which the fields $c,\psi$ vary
rapidly over distances ${\cal O}(\xi)$. 
It is then useful to define Green's
functions $G^{\pm}(\vec r,\vec r\,')$ in the regions $-\zeta <u <0$
($G^{-}$) and in $0 <u <\zeta$ ($G^{+}$) obeying
\be
\label{eq:green}
\nabla'^{2}G^{\pm}(\vec r,\vec r\,') = \delta(\vec r-\vec r\,')
\ee
and satisfying the boundary conditions, $G(\vec{r},\vec{r}')=0$ at 
$u=0$ and $u'=0$, $\partial G(\vec{r},\vec{r}')/\partial u =0$ 
at $u=\pm\zeta$, $\partial G(\vec{r},\vec{r}')/\partial u' =0$ 
at $u'=\pm\zeta$, and periodic in the other 
directions.  Note that {\it both} $\vec r$ and $\vec r\,'$ lie in the 
same region, $V_{+}$ or $V_{-}$.

Multiplying Eq. (\ref{eq:start2a}) by $G^{\pm}$ and integrating over $\vec
r\,'\in V_{\pm}$ gives
\ben
\label{eq:greqc}
\delta\mu(\vec r) &=& \int_{V_{\pm}}d\vec r\,'\frac{G^{\pm}(\vec r,\vec
r\,')}{\Gamma_{c}}\pqpq{c'}{t}   \cr
&+&\oint_{S_{\pm}}d\vec S'\cdot\Bigl(\delta\mu'\vec{\nabla}'G^{\pm} -
G^{\pm}\vec{\nabla}'\delta\mu'\Bigr) 
\een
where $\delta\mu' = \delta\mu(\vec r\,')$ defined as in
Eq. (\ref{eq:deltamu}), $\int_{V_{+}}d\vec r\,'$ denotes integration over
$V_{+}$ defined by $0 < u(\vec r\,') < \zeta $ and 
$\oint_{S_{+}}d\vec S'$ denotes 
integration over the boundaries $S_{+}$ enclosing $V_{+}$ and 
similarly for $V_{-}$ enclosed by $S_{-}$.  
Multiplying Eq. (\ref{eq:greqc}) by $\partial \ci/\partial u$ where
$\ci(u)$ is determined by the solution 
of Eqs. (\ref{eq:eqpsi}) and (\ref{eq:eqc})
for a planar interface in thermal equilibrium and
integrating over $-\zeta < u < +\zeta$ gives
\ben
\label{eq:ceqg}
{\cal B} + {\cal S} &\equiv &\int_{-\zeta}^{+\zeta}du\,\frac{\partial
\ci}{\partial u}\,\delta\mu  \cr
 &=& \int^{+\zeta}_{-\zeta}du\frac{\partial \ci}{\partial
u}\left(\frac{\partial f}{\partial c} - K_{c}\nabla^{2}c -\mu_{eq}\right)
\een
where ${\cal B} = {\cal B}^{+} + {\cal B}^{-}$ and ${\cal S} = {\cal S}^{+}
+ {\cal S}^{-}$ with 
\be
\label{eq:B+-}
{\cal B}^{\pm} =
\pm\frac{1}{\Gamma_{c}}\int_{0}^{\pm\zeta}du\frac{\partial
\ci}{\partial
u}\int_{V_{\pm}}d\vec r\,'G^{\pm}(\vec r,\vec r\,')\frac{\partial c'}
{\partial t},
\ee
\be
\label{eq:S+-}
{\cal S}^{\pm} = \pm\int_{0}^{\pm\zeta}du\frac{\partial \ci}
{\partial u}
\oint_{S_{\pm}}d\vec S'\cdot\left(\delta\mu'\vec{\nabla}'G^{\pm} 
- G^{\pm}\vec{\nabla}'\delta\mu'\right). 
\ee
An analogous formula for $\psi$ is obtained by multiplying
Eq. (\ref{eq:start}) by $\partial\psii/\partial u$, with
$\psii(u)$ the solution of Eqs. (\ref{eq:eqpsi}) and (\ref{eq:eqc}),
and integrating over $-\zeta <u <+\zeta$ to obtain
\be
\label{eq:peqg}
\frac{1}{\Gamma_{\psi}}\int_{-\zeta}^{+\zeta}du
\frac{\partial\psii}{\partial u}\frac{\partial\psi}{\partial t} 
 =  
+\int_{-\zeta}^{+\zeta}du\frac
{\partial\psii}{\partial u}\left(K_{\psi}\nabla^{2}\psi -
\frac{\partial f}{\partial\psi}\right).
\ee
Each term in the above equations can be systematically expanded in 
powers of $\epsilon$.  In this paper, attention is restricted to
the terms ${\cal O}(\epsilon)$ as much of the relevant physics is apparent
at this order.  Going to higher order in $\epsilon$ does not yield 
any new physical insight but does require considerable book-keeping 
skill.

To facilitate the expansion, $c(\vec r,t)$, $\psi(\vec r,t)$ 
and the chemical potential $\mu(\vec r,t)$ are expanded in 
a power series in $\epsilon$, 
\ben
\label{eq:expand-psic}
c(\vec r,t) &=& \ci(u(\vec r)) +\epsilon\cin{1} +
\epsilon^{2}\cin{2} + \cdots  \cr
\psi(\vec r,t) &=& \psii(u(\vec r)) + \epsilon\psiin{1} +
\epsilon^{2}\psiin{2} + \cdots  \cr
\mu(\vec r,t) &=& \mui(u(\vec r)) + \epsilon\muin{1} +
\epsilon^{2}\muin{2} + \cdots
\een
where the superscript `in' refers to the {\it inner} solution.
To expand the Laplacian in powers of $\epsilon$, it is useful to introduce a
curvilinear coordinate system with one coordinate $u$ along the local
normal to the interface and $(d-1)$ coordinates $\vec s$
perpendicular to $u$ and tangent to the interface.
For simplicity a two-dimensional system is considered where  
$s$ is the scalar arc length. Note that the 
${\cal O}(\epsilon^{0})$ terms,
$\ci$ and $\psii$ in Eq. (\ref{eq:expand-psic}), are the
equilibrium planar interface solutions of Eq. (\ref{eq:eqpsi}) 
and Eq. (\ref{eq:eqc}).

	At this early stage of the calculations it is worth pointing 
out that the exact position of the interface has not been specified. 
The choice of the exact interface position is somewhat flexible to 
within a distance $\xi$ and there is no particular reason for
choosing the interface position to be defined by $\psi(u(\vec r,t)=0) =0$
as is often done in the literature. Indeed, this particular
choice can lead to unreasonable constraints on the free energy for a
mapping between the phase field model and the sharp interface limit to be
possible.  In this work the interface will be chosen to be the Gibbs surface
defined so that the excess surface concentration is equal on both sides of
the interface. This is in essence a solvability condition which ensures that 
the chemical potential is continuous across the interface.

The transformation from Cartesian to curvilinear
coordinates (see Appendix A) leads to
the formal expansion
\be
\xi^2 \nabla^2 = \call{0} + \ep\call{1}+\ep^2\call{2}+ \cdots
\ee
where the specific form of  $\call{n}$ depends on the expansion.
In the inner region, derivatives of the fields with respect to $u$
are much larger than derivatives with respect to $s$
which are identically zero when the curvature $\kappa$ and the
P\'{e}clet number vanish.  This is taken into account by introducing
the dimensionless variables ${\bar{u}}$ and ${\bar{s}}$ which are ${\cal
O}(\epsilon^{0})$ by $u =\xi{\bar{u}}$ and $s =\xi{\bar{s}}/\epsilon$. As
shown in Appendix  A, this scaling leads to
\[
\begin{array}{l}
            \call{0} = \partial_{\us\us} \\
                \call{1} = \Curv \partial_\us \\
                \call{2} = \partial_{\scs\scs} - \Curv^2 \bar{u} \partial_{\us}
\end{array}
\]
where the dimensionless curvature, $\Curv
\equiv \xi \kappa/\epsilon $, is of order unity. 

	Lastly, the time derivatives in Eqs.\ (\ref{eq:B+-}) and 
(\ref{eq:peqg}) must be expanded in $\ep$.  For these calculations,
it is convenient to work in the frame in which the interface is stationary
so that
\be
\frac{\partial}{\partial t}\Big|_{\vec r} = \frac{\partial}{\partial
t}\Big|_{(u,s)}-\vec{v}\cdot\vec{\nabla}
\ee
where $\vec{v}$ is the interface velocity which has components normal 
and tangential to the interface.  The time derivative on the 
right hand side corresponds to relaxational dynamics not accounted 
for by motion of the interface.   When this operator acts 
on the fields, $c$ and $\psi$, the tangential component and time 
derivative are of order $\epsilon^3$ and can be dropped\cite{velocity}.  
Thus to $\sim {\cal O} (\epsilon)$,
$\partial/\partial t|_{\vec{r}}$ becomes;
\be
\frac{\partial}{\partial t}\Big|_{\vec r} = 
-\epsilon\, \frac{v_1}{\tau} \pqpq{}{\bar{u}} + {\cal O}(\epsilon)^2 + \cdots
\ee
where the normal velocity has been expanded in a power series in 
$\epsilon$,  
\be
v_{\rm normal} \equiv -\pqpq{u}{t} \equiv \frac{\xi}{\tau}\sum_{m=1}^{\infty}
\epsilon^m v_m.
\ee

Using these expansions and expanding $f$ around  
 $\ci$ and $\psii$ the right hand sides of
Eqs.\ (\ref{eq:ceqg}) and (\ref{eq:peqg}) become
\ben
&&\int_{-\zeta}^{+\zeta} du \pqpq{\ci}{u}
\Bigl[\mu_{0} + \scaleK_c \xi^2 \nabla^2 c - \pqpq{f}{c}\Bigr]
=
\nonumber \\
&=&
\int_{-\bar{\zeta}}^{+\bar{\zeta}} d\us \pqpq{\ci}{\us}
 \Big[
\left(\mu_{0}+
\scaleK_c \call{0} \ci - \ffi{1}{0}\right) +
\ep \Bigl(\scaleK_c(\call{1}\ci
\nonumber \\
 \ \ \ &+& \
\call{0}\cin{1}) -\cin{1}\ffi{2}{0}
-\psiin{1}\ffi{1}{1} \Bigr) + {\cal O}(\ep^2) \Big]
\een
and
\ben
&&\int_{-\zeta}^{+\zeta} du \pqpq{\psii}{u}
\left[\scaleK_{\psi} \xi^2 \nabla^2 \psi - \pqpq{f}{\psi}\right]
= 
\nonumber \\
&=&
\int_{-\bar{\zeta}}^{+\bar{\zeta}} d\us \pqpq{\psii}{\us}
\Big[
\Bigl(\scaleK_{\psi} \call{0} \psii - \ffi{0}{1}\Bigr)
+ \ep \Bigl(\scaleK_{\psi} (\call{1}\psii +
\nonumber \\
&+& \call{0}\psiin{1}) -\psiin{1}\ffi{0}{2}
-\cin{1}\ffi{1}{1} \Bigr) + {\cal O}(\ep^2) \Big]
\een
where $\ffi{n}{m} \equiv \partial^{n+m} f/\partial c^n \partial 
\psi^m|_{\psii,\ci}$, $\scaleK_c \equiv K_c / \xi^2$,
$\scaleK_{\psi} \equiv K_{\psi} / \xi^2$ and $\bar{\zeta} = \zeta/\xi$.
Terms of ${\cal O}(\ep^0)$ vanish by construction. For later use, it is
convenient to perform partial integrations on combinations of terms
\cite{mathtip}

\ben
\label{eq:cex}
\int_{-\bar{\zeta}}^{+\bar{\zeta}} &d\us& \pqpq{\ci}{\us}
\Bigr(\scaleK_c \pqpq{^2}{\us^2} - \ffi{2}{0}\Bigl)\cin{1}
=  
\nonumber \\
&=&
\int_{-\bar{\zeta}}^{+\bar{\zeta}} d\us\, \cin{1}
\left(\scaleK_c \pqpq{^2}{\us^2}-\ffi{2}{0}\right)\pqpq{\ci}{\us}
\nonumber \\
&=&
\int_{-\bar{\zeta}}^{+\bar{\zeta}} d\us\, \cin{1} \ffi{1}{1}
\pqpq{\psii}{\us}
\een
and
\ben
\label{eq:pex}
\int_{-\bar{\zeta}}^{+\bar{\zeta}} &d\us& \pqpq{\psii}{\us}
\Bigl(\scaleK_{\psi} \pqpq{^2}{\us^2} - \ffi{0}{2}\Bigr)\psiin{1}
=  
\nonumber \\
&=&
\int_{-\bar{\zeta}}^{+\bar{\zeta}} d\us\, \psiin{1}
\left(\scaleK_{\psi} \pqpq{^2}{\us^2}-\ffi{0}{2}\right)\pqpq{\psii}{\us}
\nonumber \\
&=&
\int_{-\bar{\zeta}}^{+\bar{\zeta}} d\us\, \psiin{1} \ffi{1}{1}
\pqpq{\ci}{\us}
\een
since derivatives of $\ci$ and $\psii$ vanish at
$\us=\pm \bar{\zeta}$ in the limit $\bar{\zeta} = \zeta/\xi >> 1$.

	To complete the calculation, the left-hand
sides of Eqs.\ (\ref{eq:ceqg}) and (\ref{eq:peqg}) are 
expanded to lowest order in $\ep$.  The 
expansion for $\psi$ in Eq. (\ref{eq:peqg}) is straightforward
\ben
\frac{1}{\gpsi}\int_{-\zeta}^{+\zeta}du\goldp\frac{\partial\psi}{\partial
t} 
&=& -\frac{\ep}{\gpsi\tau}\int_{-\bar{\zeta}}^{+\bar{\zeta}}d\us
\goldpus v_{1}\goldpus
\cr
&=& -\epsilon\frac{v_{1}\xi}{\tau}\frac{\sigma_{\psi}}{K_{\psi}\Gamma_{\psi}} 
+{\cal O}(\epsilon^{2})
\een  
  where
 \be
 \sigma_{\psi} \equiv K_{\psi} \int_{-\zeta}^{\zeta} du
\left(\goldp\right)^2.
 \ee

The equivalent expansion for $c$ is more complicated.  The algebra
is given in Appendix B. 
Formally, the results of these calculations can be written as
\ben
 {\cal B} + {\cal S} = 
 \ep ({\cal B}_1/\tau + {\cal S}_1) 
+ \ep^2 ({\cal B}_2/\tau + {\cal S}_2) + \cdots 
\een
where ${\cal B}_n$ and ${\cal S}_n$ are given in Appendix B.  

	Putting all these results together gives the
following two equations to ${\cal O}(\ep)$
\be
\label{eq:finpsi}
\frac{v_{1}\xi}{\tau}\frac{\sigma_{\psi}}{\Gamma_{\psi}K_{\psi}} =
-\frac{\bar{\kappa}\sigma_{\psi}}{\xi} - A_{1}
\ee
and
\ben
\label{eq:finc}
&&\Delta c\,\muin{1}(0,s) = -\frac{\sigma_{c}\bar{\kappa}}{\xi} + A_{1} \cr
&&-\frac{v_{1}(s)\xi^{2}}{\tau\Gamma_{c}}\int_{-\bar{\zeta}}^{+\bar{\zeta}}
d\us[\co(\us) - \ci(\us)]^{2} \cr
&&-\frac{\partial\muin{1}}{\partial\us}\Big|_{-\bar{\zeta}}
\int_{-\bar{\zeta}}^{0}d\us[\co(\us) - \ci(\us)] \cr
&&-\frac{\partial\muin{1}}{\partial\us}\Big|_{+\bar{\zeta}}
\int_{0}^{+\bar{\zeta}}d\us[\co(\us) - \ci(\us)]
\een
where $\Delta c\equiv \ci(\bar{\zeta}) - \ci(-\bar{\zeta})$
is the miscibility gap,
\be
\label{eq:couteq}
\co(\us) \equiv \left\{
\begin{array}{cc}
\ci(-\bar{\zeta}) & \ \ \ \us < 0 \cr
\ci(+\bar{\zeta}) & \ \ \ \us > 0
\end{array}
\right . 
\ee
\be
\sigma_c \equiv  K_c \int_{-\zeta}^{\zeta} du
\left(\pqpq{\ci}{u}\right)^2
\ee

and 
\be
\label{eq:a1}
A_{1} = \int_{-\bar{\zeta}}^{+\bar{\zeta}} d\us
\left(\psiin{1}\pqpq{\ci}{\us} -
\cin{1}\pqpq{\psii}{\us}\right)\ffi{1}{1} 
\ee

	Equation (\ref{eq:finc}) gives the chemical potential 
$\mu$ of the {\it inner} solution at the interface (i.e., at 
$\us=0$) which must be matched to
the {\it outer} solution at $\us = \pm\bar{\zeta}$. 
An expression for
$\muin{1}(\pm\bar{\zeta})$ can be obtained by 
expanding Eq. (\ref{eq:start2a})
to ${\cal O}(\ep)$ 
\be
\label{eq:cdeltamu}
v_1 \pqpq{\ci}{\us} =
-\frac{\gc\tau}{\xi^2}{\cal L}_{0}\,\muin{1}
\ee

Integrating Eq. (\ref{eq:cdeltamu}) twice, first from $\us$ to $\bar{\zeta}$ and
then from $0$ to $\bar{\zeta}$ yields
\ben
\label{eq:deltamu1}
\muin{1}(\bar{\zeta},s) &=& \muin{1}(0,s) 
+ \bar{\zeta}\frac{\partial\muin{1}}{\partial\us}\Big|_{+\bar{\zeta}}
\cr
&+& \frac{v_{1}\xi^{2}}{\tau\Gamma_{c}}\int_{0}^{\bar{\zeta}}
d\us[\co(\us) - \ci(\us)]
\een
and similarly for $\muin{1}(-\bar{\zeta})$. From Eqs. (\ref{eq:finc}) 
and (\ref{eq:deltamu1}) we obtain
\ben
\label{eq:mubarzeta}
&&\Delta c\muin{1}(\pm\bar{\zeta},s) = -\sigma_{c}\kappa/\ep + A_{1}
\pm\Delta c\bar{\zeta}\frac{\partial\muin{1}}{\partial\us}\Big|_{\pm
\bar{\zeta}} \cr
&&-\frac{v_{1}\xi^{2}}{\tau\Gamma_{c}}\int_{-\bar{\zeta}}^{+\bar{\zeta}}
d\us[\co(\us) - \ci(\us)]^{2} \cr
&&-\frac{\partial\muin{1}}{\partial\us}\Big|_{-\bar{\zeta}}
\int_{-\bar{\zeta}}^{0}d\us[\co(\us) - \ci(\us)] \cr
&&-\frac{\partial\muin{1}}{\partial\us}\Big|_{+\bar{\zeta}}
\int_{0}^{+\bar{\zeta}}d\us[\co(\us) - \ci(\us)] \cr
&&+\Delta c\frac{v_{1}\xi^{2}}{\tau\Gamma_{c}}\int_{0}^{\pm\bar{\zeta}}d\us
[\co(\us) - \ci(\us)]
\een
The integrals in Eq. (\ref{eq:mubarzeta}) can be written in a more useful form
by noting that $\bar{\zeta}\gg 1$ in the inner region so that $\ci(\pm\us)
= \ci(\pm\infty)$ for $|\us|\geq\bar{\zeta}$. Eq. (\ref{eq:mubarzeta}) becomes
\ben
\label{eq:deltamufinal}
&&\Delta c\muin{1}(\pm\bar{\zeta},\bar{s}) = 
-\frac{\sigma_{c}\bar{\kappa}}{\xi} +
A_{1} \pm\Delta c\bar{\zeta}\frac{\partial\muin{1}}{\partial\bar{u}}
\Big|_{\pm\bar{\zeta}} \cr
&& -\frac{v_{1}(\bar{s})\xi^{2}}{\tau\Gamma_{c}}\int_{-\infty}^{+\infty}d\us[
\ci(\us) - \co(\us)]^{2} \cr
&& -\frac{\partial\muin{1}}{\partial\us}\Big|_{-\bar{\zeta}}
\int_{-\infty}^{0}d\us[\co(\us) - \ci(\us)] \cr
&& -\frac{\partial\muin{1}}{\partial\us}\Big|_{+\bar{\zeta}}
\int_{0}^{+\infty}d\us[\co(\us) - \ci(\us)] \cr
&& +\Delta c\frac{v_{1}(\bar{s})\xi^{2}}{\tau\Gamma_{c}}\int_{0}^{\pm\infty}d\us
[\co(\us) - \ci(\us)]
\een 
One last result will be needed and is obtained by integrating Eq.
(\ref{eq:deltamu}) over $-\bar{\zeta} < \us < +\bar{\zeta}$
\be
\label{eq:vs}
v_{1} = -\frac{\tau\Gamma_{c}}{\Delta c\xi^{2}}\Bigl(\frac{\partial\delta
\mu_{1}^{\rm in}}{\partial\us}\Big|_{+\bar{\zeta}} - \frac{\partial\muin{1}}
{\partial\us}\Big|_{-\bar{\zeta}}\Bigr)
\ee
The solution for $\muin{1}(\pm\bar{\zeta})$ must be matched to the
solution in the {\it outer} region.

\subsection{Outer Expansion}
\label{sec:B}

	Far from the interface, the fields $\psi$
and $c$ vary slowly in space and are close to the bulk equilibrium 
values $\psi_{eq}$ and $c_{eq}$. Variations of the fields in the bulk regions
far from the interface take place on length scales ${\cal O}(\xi/\ep)$ in all
directions which implies that suitable dimensionless space and time coordinates
are $(\tilde{u},\tilde{s},\tilde{t}) \equiv (\ep u/\xi ,\ep s/\xi
,\ep^{2}t/\tau)$.
 
Expanding $\psi(\vec r)$ about the bulk equilibrium
solution $\psi(\vec{r}) = \delta \psiob(\vec{r})
+ \psi_{eq}$ gives
\ben 
&&\frac{\partial \delta \psiob}{\partial t} =  \gpsi
\left(K_{\psi}\nabla^2 \delta \psiob -
\frac{\partial f}{\partial \psi}\Big|_{eq} 
-  \frac{\partial^2 f}{\partial \psi^2}\Big|_{eq} \delta \psiob \right. \cr
&& \left.
- \frac{1}{2!} \frac{\partial^3 f}{\partial \psi^3}\Big|_{eq} (\delta
\psio)^2
- \frac{1}{3!} \frac{\partial^4 f}{\partial \psi^4}\Big|_{eq} (\delta \psiob)^3
- \cdots \right)
\label{eq:psiout}
\een 
By definition, $(\partial f/\partial \psi)_{eq}=0$
and, since $\delta \psiob = 0$ in the limit
$\ep \rightarrow 0$, Eq. (\ref{eq:psiout}) is linear at ${\cal O}(\ep)$.
Furthermore,
$(\partial^{2} f/\partial \psi^2)_{eq} > 0$ so that $\delta \psio$
vanishes exponentially with time for all wavelengths.  Thus, $\delta
\psio$ is trivial in the outer region and can be ignored.
It is convenient to expand $c^{\rm out}$ and $\mu^{\rm out}$ 
in the outer region as
\ben
\label{eq:outercmu}
c^{{\rm out}}(\vec r) &=& \co + \ep\con{1}  +\cdots \cr
\mu^{{\rm out}}(\vec r) &=& \muo + \ep\muon{1} + \cdots
\een
where $\co$ is given by Eq. (\ref{eq:couteq}). At ${\cal O}(\ep^{3})$
in dimensionless variables, in the lab frame
\be
\label{eq:motiondeltac1out}
\frac{\partial\con{1}}{\partial\tilde{t}} = \frac{\tau\Gamma_{c}}{\xi^{2}}
\tilde{\nabla}^{2}\muon{1}
\ee
where $\tilde{\nabla}\equiv(\xi/\ep)\nabla$ is the scaled dimensionless derivative
suitable for the outer region.
This simplifies to a linear diffusion equation for the chemical potential inside
the bulk phases which reads, in dimensional units,
\be
\label{eq:diffusionmu}
\frac{\partial\mu^{{\rm out}}}{\partial t} = D_{c}\nabla^{2}\mu^{{\rm out}}
\ee
where $D_{c}\equiv\Gamma_{c}(\partial\mu^{{\rm out}}/\partial c^{{\rm out}})_{eq}$ is a diffusion
constant. The value of $D_{c}$ depends on the bulk equilibrium phase
considered.

\subsection{Matching and the Gibbs Surface}
\label{sec:C}

\begin{figure}[btp]
\begin{minipage}{8.0cm}
\epsfxsize=8.0cm 
\epsfysize=8.0cm
\epsfbox{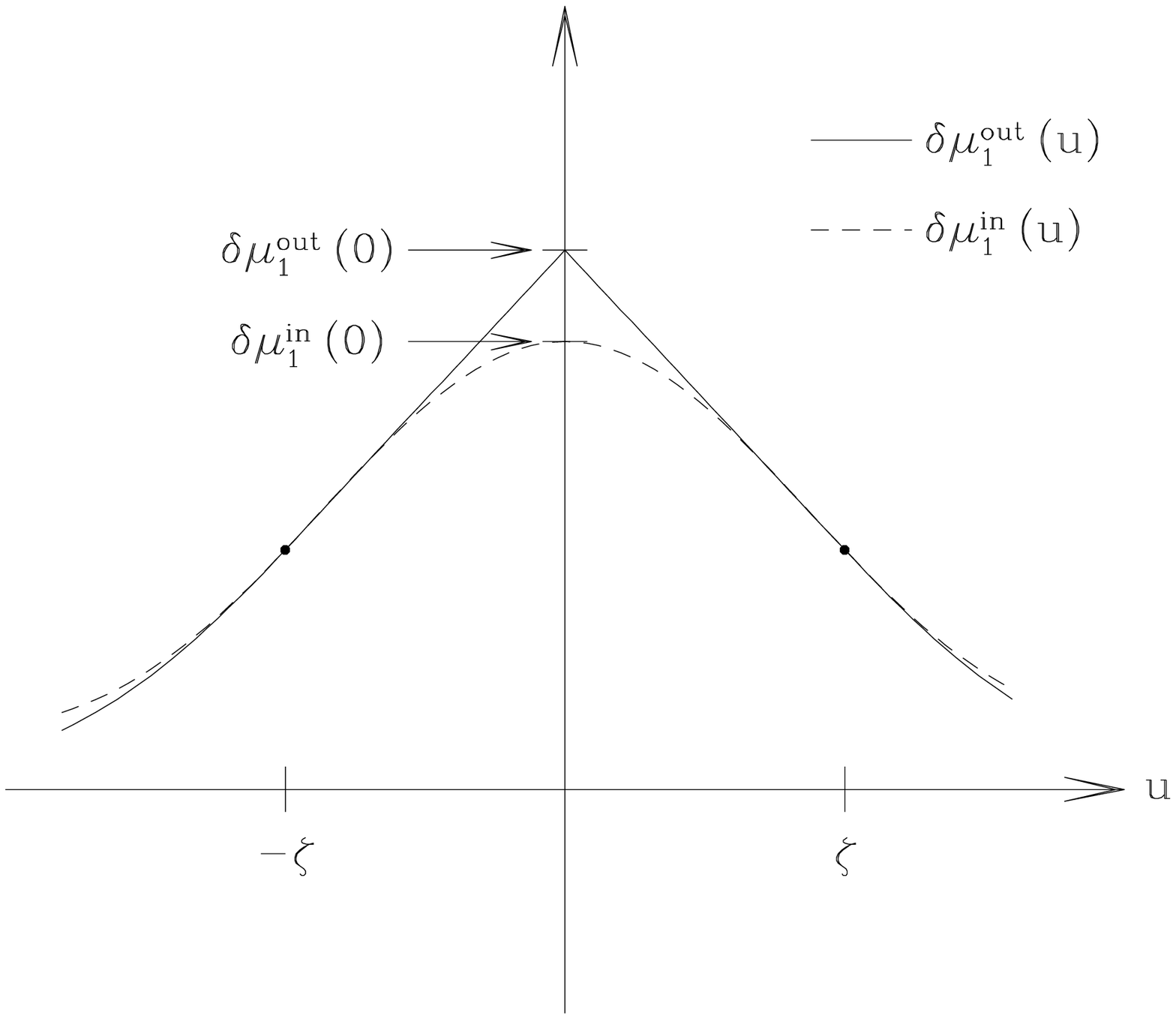}
\end{minipage}
\caption{Matching of $\muin{1}(u)$ (dashed line) with
$\muon{1}(u)$ (solid line) at $u = \pm\zeta$.}
\label{figmatch}
\end{figure}

To solve the diffusion problem of Eq. (\ref{eq:diffusionmu}), initial values
$\muon{1}(\tilde{u}=0,\tilde{s})$ are required. These are to be obtained
by matching to the inner solution $\muin{1}(\us,\scs)$ at $\us =
\bar{\zeta}$. To obtain $\muon{1}(\tilde{u}=0^{\pm},\tilde{s})$ from
$\muon{1}(\tilde{u}=\pm\tilde{\zeta},\tilde{s})$ with $\tilde{\zeta}\equiv
\ep\bar\zeta$, it is useful to Taylor expand about $\tilde{u} = \tilde{\zeta}$
\be
\label{eq:extrapmu1}
\muon{1}(\tilde{u},\tilde{s}) = \muon{1}(\pm\tilde{\zeta},
\tilde{s}) + (\tilde{u}\mp\tilde{\zeta})\frac{\partial\muon{1}}
{\partial\tilde{u}}|_{\pm\tilde{\zeta}} + \cdots
\ee
In the outer region, $\tilde{\zeta}\ll 1$ and this expansion is valid at
$\tilde{u} = 0$
\be
\label{eq:extrapmu2}
\muon{1}(\pm\tilde{\zeta},\tilde{s}) = \muon{1}(0,\tilde{s})
\pm \tilde{\zeta}\frac{\partial\muon{1}}{\partial\tilde{u}}\Big|_{\pm
\tilde{\zeta}} + \cdots
\ee
Since $\muon{1}(\pm\tilde{\zeta}) = \muin{1}(\pm\bar{\zeta})$, we
can use Eq. (\ref{eq:deltamufinal}) and Eq. (\ref{eq:extrapmu2}) to obtain
\ben
\label{eq:deltamuoutf}
&&\Delta c\muon{1}(0,\tilde{s}) = 
-\frac{\sigma_{c}\bar{\kappa}}{\xi} + A_{1} \cr
&& -\frac{v_{1}(\tilde{s})\xi^{2}}{\tau\Gamma_{c}}\int_{-\infty}^{+\infty}d\us
[\ci(\us) - \co(\us)]^{2} \cr
&& -\frac{\partial\muin{1}}{\partial\us}\Big|_{-\bar{\zeta}}\int_{-\infty}
^{0}d\us[\co(\us) - \ci(\us)] \cr
&& -\frac{\partial\muin{1}}{\partial\us}\Big|_{+\bar{\zeta}}\int_{0}^
{+\infty}d\us[\co(\us) - \ci(\us)] \cr
&& +\Delta c\frac{v_{1}(\tilde{s})\xi^{2}}{\tau\Gamma_{c}}\int_{0}^{\pm\infty}
d\us[\co(\us) - \ci(\us)]
\een
which gives the appropriate boundary value of $\muon{1}(0,\tilde{s})$. 
The inner solution $\muin{1}(0)$ differs from the outer solution
$\muon{1}(0)$ since the matching is done at $\tilde{u} = \tilde{\zeta}$
and extrapolated linearly to $\tilde{u} = 0$ by Eq. (\ref{eq:extrapmu2}). The
extrapolation and matching of $\muin{1}$ to $\muon{1}$ is
illustrated pictorially in Fig. (\ref{figmatch}).

\begin{figure}[btp]
\begin{minipage}{8.0cm}
\epsfxsize=8.0cm  \epsfysize=8.0cm
\epsfbox{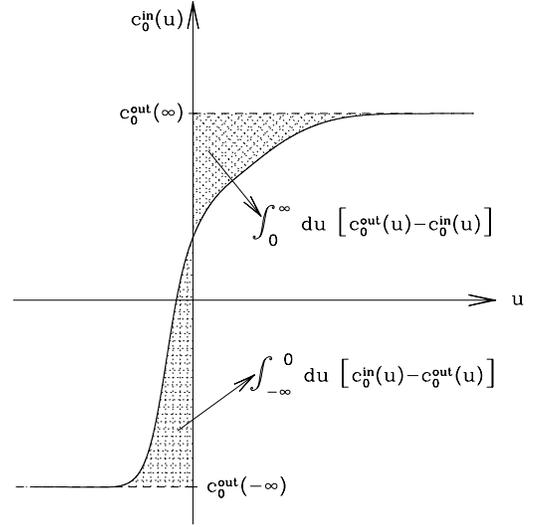}
\end{minipage}
\caption{The Gibbs surface at $u = 0$ defined by Eq. (\ref{eq:gibbs}) which
matches the concentration deficit on one side with the concentration excess on
the other.}
\label{figgibbs}
\end{figure}

The right hand side of Eq. (\ref{eq:deltamuoutf}) appears to depend on whether
the inner and outer solutions are matched at $u = +\zeta$ or $u = -\zeta$
because of the last term. This ambiguity is eliminated by defining the 
interface to be a Gibbs surface at $u = 0$ defined by the condition
\be
\int_{-\infty}^{+\infty}du[\co(u) - \ci(u)] = 0
\label{eq:gibbs}
\ee
This can always be satisfied by choosing the position of the interface at $u =
0$ to be such that Eq. (\ref{eq:gibbs}) is satisfied. In essence, the interface
position is determined by the condition that the excess concentration on one
side of the interface is exactly compensated by the deficit on the other, as
shown in Fig. (\ref{figgibbs}). This can be regarded as a solvability condition.

Another result that will be needed is obtained by matching the first 
derivative of
the chemical potential is obtained from Eq. (\ref{eq:vs}) which reads, in
dimensional units
\be
\label{eq:vs1}
v\Delta c = -\Gamma_{c}\Bigl(\frac{\partial\delta\mu^{{\rm in}}}{\partial u}\Big|_{+\zeta} -
\frac{\partial\delta\mu^{{\rm in}}}{\partial u}\Big|_{-\zeta}\Bigr).
\ee
Matching derivatives of the inner and outer solutions
for $\mu$ and extrapolating back 
to $u = 0^{\pm}$ by Eq. (\ref{eq:extrapmu1}),
gives the standard result
\be
\label{eq:vs2}
v\Delta c = -\Gamma_{c}\Bigl(
\frac{\partial\mu^{\rm out}}{\partial u}\Big|_{0^{+}} -
\frac{\partial\mu^{\rm out}}{\partial u}\Big|_{0^{-}}\Bigr)
\ee
since $\mu^{\rm out}(u)$ is linear for $0 <|u| \leq \zeta$ from 
Eq. (\ref{eq:extrapmu1}).
Finally, combining Eqs. (\ref{eq:deltamuoutf}), (\ref{eq:gibbs}) and 
(\ref{eq:vs2}),
gives the chemical potential at a moving, curved interface 
\be
\label{eq:gt1}
\Delta c(\mu^{\rm out}(0,s) - \mu_{eq}) = -\sigma_{c}\kappa + {\cal E}^2v 
+ A_{1} + {\cal O}
(\ep^{2})
\ee
where $A_{1}$ is given by Eq. (\ref{eq:a1}) and   
\be
\label{eq:kinetic1}
{\cal E}^2\equiv\frac{1}{\Gamma_{c}}\int_{-\infty}^{+\infty}du\Bigl([\co(u)]^{2}
-[\ci(u)]^{2}\Bigr).
\ee

\section{Summary of Results}

	All the results can be combined into a single set of 
boundary layer equations that can describe many different 
physical phenomena.  Typically the boundary layer equations are 
written in terms of the concentration, which is the outer 
region is simply related (to order $\epsilon$) to the chemical 
potential by the relationship 
\be
\delta \mu = \left.\pqpq{\mu}{c}\right|_{eq}\delta c.
\ee
Combining this result with Eqs. (\ref{eq:gt1}) and 
(\ref{eq:finpsi}) gives the 
Gibbs Thomson relation in dimensionless units 
\be
\label{eq:gtall}
\frac{\delta c(0,s)}{\Delta c} 
= -d_o \kappa(s) - \beta v.
\ee
where $d_o$ is the capillary length given by,
\be
d_o = \frac{\sigma}{(\Delta c)^2 (\partial \mu/\partial c)_{eq}},
\ee
$\sigma \equiv \sigma_c+\sigma_\psi$ is the total surface tension given 
by,
\be
\sigma=\int_{-\infty}^{\infty} du\left[ K_\psi \left(\pqpq{\psii}{u}\right)^2
+K_c\left(\pqpq{\ci}{u}\right)^2\right],
\ee
and $\beta$ is the coefficient of kinetic undercooling given by
\be
\beta = \frac{1}{(\Delta c)^2(\partial \mu/\partial c)_{eq})}
\left[\frac{\sigma_\psi}{K_\psi \Gamma_\psi} - {\cal E}^2\right].
\ee
Equation (\ref{eq:gtall}) provides a boundary condition 
at the interface for the diffusion equation of Eq. (\ref{eq:diffusionmu})
which can be written as
\be
\label{eq:diffcall}
\pqpq{\delta c}{t} = D_c\nabla^2 \delta c,
\ee
where
\be
D_c \equiv \Gamma_c \left.\pqpq{\mu}{c}\right|_{eq}.
\ee
This must be solved in conjunction with Eq. (\ref{eq:vs2}) which may 
be written,
\be
\label{eq:velall}
\Delta c\, v(s) = 
\left[D_c\pqpq{\delta c}{u}\right]_{0^-}
-\left[D_c\pqpq{\delta c}{u}\right]_{0^+}.
\ee

	To understand of the significance of each term 
that enters the above equations it is useful to consider 
some limiting cases.  First consider the case in which 
the concentration field is a constant slightly different 
from $c_{eq}$, the equilibrium value, $c = c_{eq} +\delta c$.
If $\delta \mu$ is the 
chemical potential difference between the phases defined by 
the non-conserved field at $u=+\infty$ and $-\infty$, 
then Eq. (\ref{eq:gtall}) reduces to the Allen-Cahn equation 
in a field,
\be
\label{eq:acH}
v = -K_\psi \Gamma_\psi \left(\kappa 
+\frac{\delta \mu\, \Delta c}{\sigma_\psi}\right).
\ee
From this point of view the kinetic undercooling can be thought of 
as the simply the relaxation of surface tension in a non-conserved 
field. Thus the Gibbs Thomson equation is equivalent to the 
Allen Cahn equation in the appropriate limit.

	The other simplifying case is when the non-conserved 
field is a constant as in a pure liquid or solid phase.  
In this case, the sharp interface equations remain the same except 
the coefficients $\sigma$ and $\beta$ become
\be
\label{eq:sigc}
\sigma = \sigma_c
\ee
and 
\be
\label{eq:betac}
\beta = 
- ({\cal E}/\Delta c)^2/(\partial \mu/\partial c)_{eq}.
\ee
For the conserved case of Model B, $\beta$ is always negative.
This term takes into account the lag of the concentration field 
behind a moving front.  When the interface is moving, the 
interfacial profile cannot instantaneously relax to the correct 
equilibrium shape $\ci$.  For the conserved field, this correction 
is roughly as important as dynamic relaxation in the bulk phase, as 
will be seen in the next section.

\section{Linear Analysis} 

\begin{figure}[btp]
\begin{minipage}{8.0cm}
\epsfxsize=8.0cm  \epsfysize=8.0cm
\epsfbox{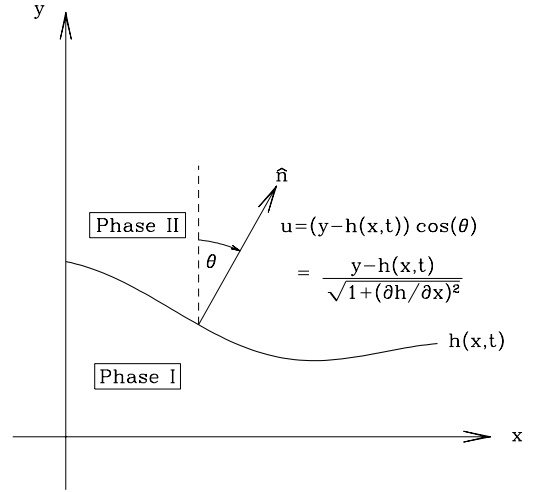}
\end{minipage}
\caption{Interface in Cartesian coordinates}
\label{figkpz}
\end{figure}

	To illustrate the influence of the various terms that 
enter the sharp interface model it interesting to study the 
dynamics of fluctuations around an almost planar interface. 
To fix ideas it is useful to consider an interface separating 
two phases that is defined by the equation $y=h(x,t)$
as shown in Fig. (\ref{figkpz}).  In the calculations to 
follow it will be assumed that $\vec{\nabla} h(x,t)$ is 
a small parameter.  This is an additional constraint not 
implicit in the sharp interface models.

To facilitate the analysis it is worth noting that the normal 
velocity and curvature can be written in terms of derivatives 
of $h$ as follows:
\be
\label{eq:vcart}
v = -\pqpq{u}{t} = \frac{1}{\gamma} \pqpq{h}{t} 
\ee
and
\be
\label{eq:kcart}
\kappa = -\frac{1}{\gamma^3}\pqpq{^2h}{x^2},
\ee
where $\gamma \equiv \sqrt{1+(\partial h/\partial x)^2}$.

\subsection{Non-conserved dynamics}

As discussed in the preceding section, the sharp interface equations 
reduce to the Allen-Cahn equation in a field when the conserved 
field is a constant.  When a single valued interface as described 
above is considered, Eq. (\ref{eq:acH}) reduces 
to the Kardar, Parisi, Zhang (KPZ) equation \cite{KPZ} in 
the absence of noise.  Substituting Eqs. (\ref{eq:vcart}) and 
(\ref{eq:kcart}) into Eq. (\ref{eq:acH}) and linearizing 
in $h$ gives

\be
\label{eq:kpz}
\frac{\partial h}{\partial t} = \nu \frac{\partial^{2}h}{\partial x^{2}} +
\frac{\lambda}{2}\Bigl(\frac{\partial h}{\partial x}\Bigr)^{2}
\ee
where $h\rightarrow  h -\lambda t$, $\nu \equiv \Gamma_{\psi}K_{\psi}$ 
and $\lambda\equiv -\nu\,\delta \mu\,\Delta c/\sigma_\psi$. The additive noise 
term $\eta(x,t)$ in the
standard KPZ equation \cite{KPZ} appears when stochastic noise 
is included in the fundamental Langevin equations.

As a specific example, consider the following free energy:
\ben
{\cal F} &=& \int d\vec{r}\left( -\frac{a}{2}\psi^2
+\frac{K_\psi}{2}|\vec{\nabla}\psi|^2\right. \nline
&&\left.+ \frac{b}{4}\psi^4
+ d\, \delta c\, \psi + \frac{w}{2}(\delta c)^2\right)
\een
where $\delta c \equiv c - c_{eq}$ and $c$ is a constant.  For this 
free energy, a planar interface is stationary when $d \rightarrow 0$.  
This interface is defined by the equations:
\ben
\psii &=& \psi_{eq} \tanh\left(\frac{u}{2\xi}\right), \nline
\ci &=&c_{eq} - \frac{d}{w} \psii, \nline
\mu_{eq} &=&0,
\een
where $\xi = \sqrt{K_\psi/2 a}$ and $\psi_{eq}=\sqrt{a/b}$.  
Thus the miscibility gap, surface tension, and 
deviations of the chemical potential are given by
\be
\Delta c = -\frac{d}{w}\Delta \psi = -2\frac{d}{w}\psi_{eq},
\ee
\be
\sigma_\psi = \frac{2}{3}\frac{K_\psi \psi^2_{eq}}{\xi}, 
\ee
and 
\be
\delta \mu = \pqpq{\mu}{c}\delta c = w \delta c.
\ee
Thus the coefficient $\lambda$ is given as
\ben
\lambda &=& \Gamma_\psi K_\psi (d\delta c) \frac{\Delta \psi}{\sigma_\psi} 
= \Gamma_\psi(d\delta c) \frac{3}{a}\sqrt{\frac{K_\psi b}{2}}.
\een

\subsection{Conserved dynamics}

	Now consider an almost planar interface separating two phases 
of different concentration with the same free energy as occurs for example
in spinodal decomposition.   Since concentration is a conserved field 
Eqs. (\ref{eq:gtall}), (\ref{eq:diffcall}), and (\ref{eq:velall}) must 
be solved simultaneously.  For simplicity, a two sided model  
in which $\partial\mu/\partial c$ is the same on both sides of the interface 
will be considered. This implies that the parameters $d_{o}$,
$\beta$ and $D$ are the same in both phases. It is straightforward to
perform the calculations in the more general 
case but this does not introduce any
new physics and is not very illuminating.  
In the limit $(\partial h/
\partial x)^{2}\ll 1$ it is 
convenient to seek solutions of Eq. (\ref{eq:diffcall})
of the form
\be
\label{eq:deltacb}
\delta c(u,x) = \delta c(0)e^{(ikx - q|u| + \omega t)},
\ee
where $u\approx y-h(x,t)$.
For this perturbation it is easy to show that the 
dimensionless quantities $\bar{\omega}\equiv
\omega d_{o}^{2}/D_{c}$, $\bar{k}\equiv kd_{o}$, $\bar{q}\equiv qd_{o}$ and the
dimensionless kinetic coefficient $\bar{\beta}\equiv \beta D_{c}/d_{o}$ 
satisfy
\ben
\label{eq:kineticsd}
&& \bar{\omega} = \frac{-2\bar{q}^{3}}{1-2\bar{q}(1+|\bar{\beta}|)} \cr
&& \bar{k}^{2} = \bar{q}^{2} - \bar{\omega}.
\een

\begin{figure}[btp]
\begin{minipage}{8.0cm}
\epsfxsize=8.0cm
\epsfysize=8.0cm
\epsfbox{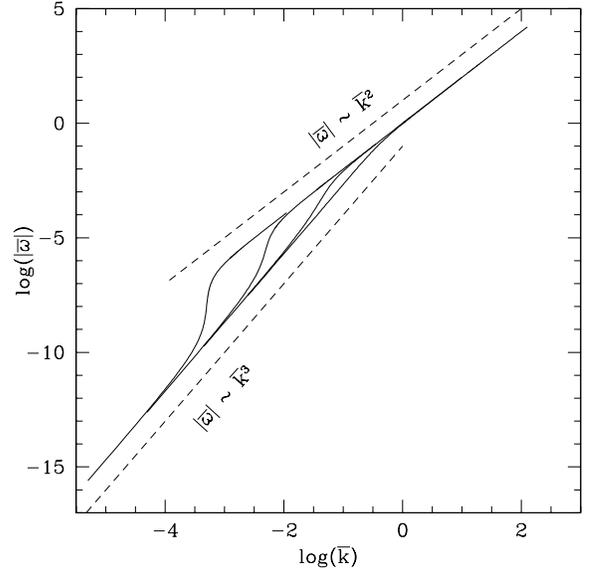}
\end{minipage}
\caption{Dispersion relation for a planar interface separating two 
conserved phases of equal free energy.  The four lines plotted in 
this figure correspond from right to 
left $-\bar{\beta} = 0, 10, 100$ and $1000$.}
\label{figsd}
\end{figure}

\noindent In the long wavelength limit ($\bar{k}\rightarrow 0$),
$\bar{k}\approx\bar{q}$ and 
$\bar{\omega} \approx -2\bar{k}^{3}$, as
expected. It is also interesting to note that in the short 
wavelength limit
($\bar{k}\rightarrow\infty$), 
$\bar{q}\rightarrow 1/(2+2|\bar{\beta}|)$ 
and $\bar{\omega} \approx -\bar{k}^2$.
The crossover from
the diffusion limited $\omega \sim -k^{2}$ at short wavelength to the asymptotic
long wavelength behavior $\omega \sim -k^{3}$ occurs at smaller values of $k$
as the kinetic coefficient $\beta$ becomes more negative. This behavior is
sketched in Fig. (\ref{figsd}).

	This analysis shows that the term $1+\bar{\beta}$ gives 
rise to diffusive (i.e., $\omega \sim -k^2$) behavior and is 
associated with relaxation in the bulk and in the interface shape. 
For example if solutions of the form $e^{ikx-q|u|}$ are 
sought instead of Eq. (\ref{eq:deltacb}) this term becomes 
simply $\bar{\beta}$.  Thus the `$1$' represents diffusive 
relaxation in the bulk and the $\bar{\beta}$ represents 
relaxation of the interface shape.

	A model commonly used to study 
spinodal decomposition is known as the Cahn-Hilliard model 
or Model B in the Halperin and Hohenberg classification 
scheme.  The free energy for this model can be written:
\be
{\cal F} = \int d\vec{r}\left(-\frac{a}{2}c^2+\frac{b}{4}c^4
+\frac{K_c}{2}|\vec{\nabla}c|^{2}\right)
\ee
and
\be
\pqpq{c}{t} = \Gamma_c \nabla^2 \fqfq{{\cal F}}{c}.
\ee
For this model a stationary planar interface is given by
\be
\ci = c_{eq}\tanh\left(\frac{y}{2\xi}\right),
\ee
where $\xi = \sqrt{K_c/2a}$ and $c_{eq}=\sqrt{a/b}$.
The coefficients entering the sharp interface equations
are then:
\be
d_o = \frac{1}{6}{\xi},
\ee
\be
\beta = -\frac{\xi}{D_c}\ \ \rightarrow \ \  \bar{\beta}=-6,
\ee
and
\be
D_c = 2 \Gamma_c a.
\ee

\subsection{Non-conserved and conserved Dynamics}

Now consider the stability of a stable phase invading a super-saturated 
liquid phase at constant velocity. This is precisely the situation
considered by Langer et al \cite{DENDRITES} in the absence 
of kinetic undercooling
and without relaxational kinetics in the bulk phases. It is easy to
show that the only solution for a planar front moving at constant velocity
which is consistent with the sharp interface model 
(i.e., Eqs. (\ref{eq:gtall}), (\ref{eq:diffcall}), and 
(\ref{eq:velall}))
is
\be
\label{eq:front1}
\frac{\delta c_0}{\Delta c} = \left\{
\begin{array}{cc}
{\rm exp}(-vy'/D_{c})-1-\beta v & y'>0 \cr
-\beta v     & y'<0
\end{array}
\right.
\ee
where $v$ is the velocity of the front and $y' = y-vt$ is a 
coordinate in the co-moving reference frame.

\begin{figure}[btp]
\begin{minipage}{8.0cm}
\epsfxsize=8.0cm
\epsfysize=8.0cm
\epsfbox{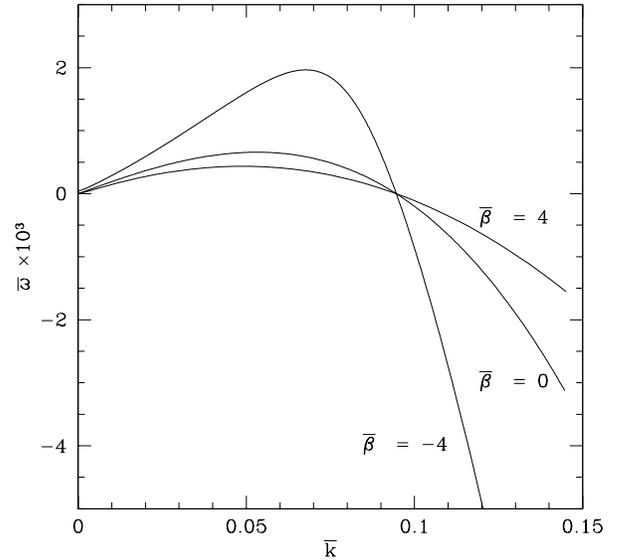}
\end{minipage}
\caption{The linear dispersion relation for the Mullins 
Sekerka instability including
the kinetic coefficient $\bar{\beta}$, for $\bar{l}=100$
}
\label{figms}
\end{figure}

The stability of this moving front can be determined by studying perturbations
about the planar front solution of Eq. (\ref{eq:front1}). We seek solutions of
the form
\be
\label{stability1}
\frac{\delta c}{\Delta c} = \frac{\delta c_{0}}{\Delta c} + \left\{
\begin{array}{cc}
\delta_{l}{\rm exp}(i\vec k\cdot\vec x + \omega t - qy') & y'>0 \cr
\delta_{s}{\rm exp}(i\vec k\cdot\vec x + \omega t + q'y') & y'<0 
\end{array}
\right.
\ee
where $q,q' >0$  and
the position of the perturbed front is at 
\be
y' = h(\vec x,t) \equiv h_{k}{\rm exp}(i\vec k\cdot\vec x + \omega t).
\label{stability2}
\ee
To linear order in $h_{k}$ and $\delta_{l,s}$, it is straightforward to show
that the dimensionless $\bar{\omega}$ and $\bar{k}$ are determined by
\ben
\label{msdispersion}
\bar{q}' &=& \bar{q} - 2/\bar{l} \cr
\bar{\omega} &=& \frac{2(\bar{q}-2/\bar{l})(1/\bar{l}-\bar{q}(\bar{q}-1/\bar{l}))}
{1-2(\bar{q}-1/\bar{l})(1-\bar{\beta})} \cr
\bar{k}^{2} &=& \bar{q}(\bar{q}-2/\bar{l}) - \bar{\omega}
\een
where $\bar{l}\equiv 2D_{c}/d_{o}v$ and $\bar{\beta}\equiv\beta D_{c}/d_{o}$. The
dispersion relation for $\bar{\omega}(\bar{k})$ is plotted in Fig. (\ref{figms}) for
$\bar{l} = 100$ and several values of $\bar{\beta}$. Note that, in contrast to Model
B where the kinetic coefficient has a definite sign $\beta < 0$, in Model C
it can have either sign. Note also that, when $\bar{l}\rightarrow\infty$,
Eq. (\ref{msdispersion}) reduces to the result of Eq. (\ref{eq:kineticsd})
for conserved Model B dynamics, as it should.

A simplified version of the general system discussed in this paper has been
extensively used to study single phase solidification phenomena
\cite{karm}. The free energy ${\cal F}$ can be written as
\be
\label{eq:freesolid}
{\cal F} = \int d\vec r\left(f(\psi) + \frac{b\lambda}{2}\Phi^{2} + \frac{W^{2}}{2}|\vec\nabla\psi|^{2}\right)
\ee
where  $\Phi\equiv c + h(\psi)$ with
\ben
\label{eq:f-h}
f(\psi) &=& -\frac{\psi^{2}}{2} + \frac{\psi^{4}}{4} \cr
h(\psi) &=& \frac{15}{16}\Bigl(\psi - \frac{2}{3}\psi^{3} + \frac{1}{5}
\psi^{5}\Bigr).
\een
With this form of $f(\psi)$, the interface width is $W$.
The dynamical evolution is governed by Langevin equations for the conserved
field $c$ and the non conserved $\psi$ which, in the noiseless limit, are
\ben
\label{eq:langevinsolid}
\frac{\partial\psi}{\partial t} &=& -\frac{1}{\tau}\frac{\delta{\cal
F}}{\delta\psi} \cr
\frac{\partial c}{\partial t} &=& \frac{D_{c}}{b\lambda}\nabla^{2}\frac{\delta
{\cal F}}{\delta c}.
\een
A stationary planar interface is given by
\ben
\label{eq:solideq}
\psii(u) &=& {\rm tanh}\Bigl(\frac{u}{W\sqrt{2}}\Bigr) \cr
\ci(u) &=& -h(\psii).
\een
For an interface with curvature $\kappa$ propagating with velocity $v$, it is
tedious but straightforward to use the Gibbs-Thomson relation of Eq.
(\ref{eq:gtall}) to find
\ben
\label{eq:gtkr}
&&\Phi(0) = -\Bigl(\frac{W}{\lambda}\Bigr)\Bigl(\frac{5}{4\sqrt{2}}
\Bigr)\kappa \cr
&&-\Bigl[\Bigl(\frac{5}{4\sqrt{2}}\Bigr)\Bigl(\frac{\tau}{W\lambda}\Bigr) -
\Bigl(\frac{209\sqrt{2}}{840}\Bigr)\Bigl(\frac{W}{D}\Bigr)\Bigr]v.
\een
The term inside the square brackets of Eq. (\ref{eq:gtkr}) is proportional to
the kinetic coefficient $\beta$ and contains the sum of the
positive Model A contribution and the negative Model B part. In principle the
kinetic coefficient $\beta$ can be of either sign in solidification
processes while it must be negative for any process described by Model B, such
as phase separation in binary alloys.

It is also trivial to show that
\ben
\label{eq:trivia}
v &=& -D_{c}\Bigl(\frac{\partial c}{\partial u}\Big|_{+} - \frac{\partial c}
{\partial u}\Big|_{-}\Bigl)  \cr
\frac{\partial c}{\partial t} &=& D_{c}\nabla^{2}c.
\een
These results are identical to those found numerically by others \cite{karm}.

\section{Summary}

The use of continuum phase-field models to describe phenomena 
involving the motion of well-defined sharp interfaces is 
discussed.   The phase-field models 
involve interfaces which are diffuse on a length scale 
of $\xi$.  Considering a general class of phase-field models, 
it is shown how equations describing the sharp-interface 
limit are obtained when $\xi\kappa \ll 1$ 
and $\xi v/D \ll 1$.  It is also shown that the 
Allen-Cahn equation is a special case of the Gibbs-Thomson 
relation.

	In particular, it should be emphasized that these 
calculations are independent of the specific form of the 
free energy functional, provided ${\cal F}$ describes 
well defined phases.  Furthermore the calculations are 
universal: a large class of free energies 
give rise to sharp interface equations which differ only in
the values of parameters but are of the same functional form.
To realize this goal, it is essential that the 
``sharp-interface limit'' involves an interface of 
finite width, $\xi$.
Expansions involving a zero width interface require 
a delicate and unphysical tuning of parameters in 
the free energy for thermodynamic consistency.  In 
this work, the small parameters $\kappa\xi$ and $v\xi/D$ 
vanish when the curvature $\kappa$ and the inverse diffusion length 
$v/D$ go to zero for a {\it finite} interface thickness $\xi$. 
Thus, delicate tuning is {\it not\/} required for thermodynamic
consistency in our approach which is based on the fundamental 
principles of statistical mechanics.

This work opens the way to construct physically consistent sharp 
interface descriptions of more complicated multiple phase systems 
such as a solid in contact with a fluid which can support flows. This 
will involve mode coupling terms in the dynamical equations. Once 
such Langevin equations are constructed, there should be no conceptual 
difficulty in deriving the corresponding interface equations.

\section*{Acknowledgments}

This work was supported by the Natural Sciences and Engineering
Research Council of Canada (MG), {\it le Fonds pour la Formation de
Chercheurs et l'Aide \`a la Recherche du Qu\'ebec\/} (MG), 
Research Corporation grant CC4787 (KRE),
NSF-DMR Grant 0076054 (KRE), and NASA Grant NAG3-1929 (JMK).

\appendix
\section{Curvilinear Coordinates}

\begin{figure}[btp]
\begin{minipage}{8.0cm}
\epsfxsize=8cm \epsfysize=8cm
\epsfbox{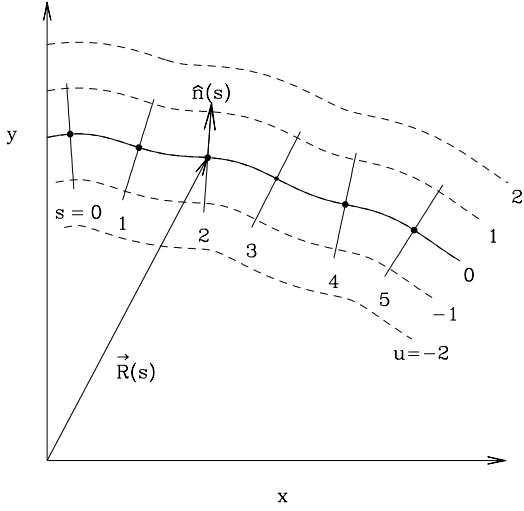}
\end{minipage}
\caption{Curvilinear Coordinates}
\label{curvcoord}
\end{figure}

	The curvilinear coordinates $(u,s)$ used in the text 
are related to the Cartesian coordinates in the following 
manner
\begin{equation}
\vec{r} = \vec{R}(s) + u \hat{n}(s),
\end{equation}
where $\vec{R}$ is the position of the interface and $\hat{n}(s)$ is the
normal vector (see Fig. \ref{curvcoord}).   The metric tensor
$g_{\alpha\beta}$ of
the transformation from Cartesian to curvilinear coordinates is
\[ g = \left( \begin{array}{cc}
1 & 0 \\
0 & (1+u\kappa)^2 
\end{array} \right) \]
where $\kappa = \partial \theta/\partial s$ is the curvature with
$\theta$ the angle between the $x$-axis and the tangent to the
curve.  The Laplacian in $(u,s)$ is then obtained in the usual 
manner
\ben
\nabla^2 &=&
\sum_{\alpha,\beta} \frac{1}{\sqrt{|g|}}\pqpq{}{x^\alpha}
\sqrt{|g|} g^{\alpha\beta} \pqpq{}{x^{\beta}}
\nonumber \\
&=&
\pqpq{^2}{u^2} + \frac{\kappa}{1+u\kappa} \pqpq{}{u}
+ \frac{1}{(1+u\kappa)^2} \pqpq{^2}{s^2} 
\nonumber \\
&& \ \ \ \ \ \ 
- \frac{u \kappa_s}{(1+u\kappa)^3} \pqpq{}{s} 
\een
where $x^1 = u$, $x^2=s$, $g^{\alpha\beta}$ are the components of
the inverse of the matrix $g$ and 
$\kappa_s \equiv \partial \kappa/\partial s$.

	In the inner region, the fields vary much more rapidly in
the $u$ direction than the $s$ direction. The
coordinates $(u,s)$ are rescaled in dimensionless units as 
$(\us,\scs) \equiv (u/\xi,\epsilon s/\xi)$.  The dimensionless curvature
$\bar{\kappa} = \xi\kappa/\epsilon$ and $\bar{\kappa}_{\bar{s}} =
\xi^{2}\kappa_{s}/\epsilon^{2}$. In terms of these dimensionless
variables,the Laplacian becomes
\ben
\xi^2 \nabla^2 = \pqpq{^2}{\us^2} &+& \frac{\ep\Curvapp}{1+\ep\us\Curvapp}
\pqpq{}{\us} +\frac{\ep^2}{(1+\ep\us\Curvapp)^2} \pqpq{^2}{\scs^2}
\nonumber  \\
&-&
\frac{\ep^3 \us\Curvapp_\scs}{(1+\ep\us \Curvapp)^3}
\pqpq{}{\scs}
\nonumber \\
= \pqpq{^2}{\us^2} &+& \Curvapp \left( \ep \sum_{n=0} (-\ep\us\Curvapp)^n
\right) \pqpq{}{\us} \nonumber  \\
&+&
\left( \ep^2 \sum_{n=0} (n+1) (-\ep\us\Curvapp)^n \right) \pqpq{^2}{\scs^2}
\nonumber  \\
&-&
  \frac{\us \Curvapp_\scs}{2}\left( \ep^3 \sum_{n=0} (n+1)(n+2)
(-\ep\us\Curvapp)^n \right)\pqpq{}{\scs} \nonumber \\
&=&
\call{0}  + \ep \call{1} + \ep^2 \call{2} + \ep^3 \call{3} + \cdots
\een
where 
\ben
\call{0} &=& \pqps{^2}{\us^2}  \\
\call{1} &=& \Curvapp \pqps{}{\us}  \\
\call{2} &=& \pqps{^2}{\scs^2} - \Curvapp^2 \us \pqps{}{\us}  \\
\call{3} &=& -2 \us \Curvapp \pqps{^2}{\scs^2} + \Curvapp^3 \us^2
\pqps{}{\us} - \us\Curvapp_{\bar{s}} \pqps{}{\scs}. \een

\section{Green's Functions}

	It will also be useful to develop an expansion 
for the inverse of the Laplacian or Green's function.
The Green's function of interest is defined by
\be
\nabla^{2}_{\vec{r}} G(\vec{r},\vec{r}\,') = \delta(\vec{r}-\vec{r}\,').
\ee
An expansion of the Green's function can be obtained in
a straightforward manner.  Let $G(\vec{r},\vec{r}\,') =
 G_0(\us,\scs;\us',\scs') +  \epsilon G_1(\us,\scs;\us',\scs') + \cdots$
where
\ben
&&\call{0} G_0 = 0 \\ 
&&\call{0} G_1 + \call{1} G_0 
= \delta(\us-\us')\delta(\scs-\scs') \\
&&\call{0} G_2 + \call{1} G_1 + \call{2} G_0 
= 0
\een
and so on.  The solution for $G_0$
is $G_0=0$, so that the lowest order solution
for $G$ is $G_1$, which satisfies the equation
\be
\label{eq:g1}
\pqpq{^2}{\us^2} G_1(\us,\scs;\us',\scs') = \delta(\us-\us')\delta(\scs-\scs') 
\ee
which has the solution
\[
G^-_1(\us,\scs;\us',\scs') =  \left\{
\begin{array}{cc}
+\us\delta(\scs-\scs') & \ \  -\zeta < \us' < \us < 0 \\
+\us'\delta(\scs-\scs') & \ \ -\zeta < \us < \us' < 0, \\
\end{array}
\right.
\]

\[ G^+_1(\us,\scs;\us',\scs') =  \left\{
\begin{array}{cc}
-\us'\delta(\scs-\scs') & \ \ \ \ 0 < \us' < \us < +\zeta  \\
-\us\delta(\scs-\scs') & \ \ \ \  0 < \us < \us'  < +\zeta. 
\end{array}
\right.
\]

The surface terms ${\cal S}^{\pm}$ of Eq. (\ref{eq:S+-}) can be
expanded as
\ben
 {\cal S}^\pm &=& \pm \int_{0}^{\pm\bar{\zeta}} d\us \goldcus
\oint_{B}d\scs' \Bigl[\delta\mu(\us',\scs') \pqpq{G^\pm}{\us'}
\nonumber 
\\
&& 
\ \ \ \ \ \ \ - 
 \ G^\pm \pqpq{\delta\mu(\us',\scs')}{\us'}
\Bigr]\Big|_{B}
\nonumber \\
&=&\pm \int_{0}^{\pm\bar{\zeta}} d\us
  \goldcus\oint_{B} d\scs' \sum_{n=1}\sum_{m=0} \ep^{n+m-1} \times
\nonumber \\
&& \ \ \ \ \times
\Bigl[\delta\mu_n(\us',\scs') \pqpq{G^\pm_m(\us,\scs;\us',\scs')}{\us'}
\nonumber \\
&& \ \ \ \ \ \ \
- \ G^\pm_m(\us,\scs;\us',\scs') \pqpq{\delta\mu_n(\us',\scs')}{\us'}
\Bigr]\Big|_{B}
\nonumber \\
&=& \ep {\cal S}^\pm_1 + \ep^2 {\cal S}^\pm_2 + \cdots
\een
where
\ben
 {\cal S}^\pm_1 &=& \pm \int_{0}^{\pm\bar{\zeta}}d\us\goldcus\oint_{B}d\scs'
\left(\delta\mu_1\pqpq{G^\pm_1}{\us'} - G^\pm_1 \pqpq{\delta\mu_1}{\us'}
\right)\Big|_{B} 
\nonumber \\
 {\cal S}^\pm_2 &=& \pm \int_{0}^{\pm\bar{\zeta}} d\us
\goldcus \oint_B d\scs'  \Bigl[
\delta\mu_2\pqpq{G^\pm_1}{\us'}
+\delta\mu_1\pqpq{G^\pm_2}{\us'}
\nonumber
\\
&& \ \ \
-  \ G^\pm_2 \pqpq{\delta\mu_1}{\us'}
- G^\pm_1 \pqpq{\delta\mu_2}{\us'}
\Bigr]\Big|_{B} 
\een
where the subscript $B$ indicates that the integrands are evaluated on the
boundary at $\us' = 0^{\pm}$ and at $\us' = \pm\bar{\zeta}$.

	The ${\cal O}(\epsilon)$ surface contribution becomes

\ben
{\cal S}_{1} &=& {\cal S}_{1}^{-} + {\cal S}_{1}^{+} \cr
&=& \int_{-\bar{\zeta}}^{0} d\us \goldcus\left(\delta\mu_{1}^{\rm in}(0,\scs) +
\us\frac{\partial\delta\mu^{\rm in}_1(\us',\scs)}{\partial\us'}\Big|_{-\bar\zeta}
\right) \nonumber \\
&+& \int_0^{+\bar{\zeta}} d\us \goldcus
\left(\us\frac{\partial\delta\mu_1^{\rm in}(\us',\scs)}{\partial\us'}\Big|_{+\bar
{\zeta}}+\muin{1}(o,\scs)\right) \nonumber \\
&=& \delta\mu_1^{\rm in}(0,\scs) \nonumber \\
&+& \frac{\partial\delta\mu^{\rm in}_1}{\partial\us'}\Big|_{-\bar{\zeta}}
\int_{-\bar{\zeta}}^{0}d\us[\ci(-\bar{\zeta}) - \ci(\us)] \nonumber
\\
&+& \frac{\partial\delta\mu_1^{\rm in}}{\partial\us'}\Big|_{+\bar{\zeta}}
\int_{0}^{+\bar{\zeta}}d\us[\ci(+\bar{\zeta}) - \ci(\us)].
\label{eq:BS}
\een

To evaluate the bulk contribution ${\cal B}$,
Eq. (\ref{eq:B+-}) is expanded in powers of $\epsilon$. Eq. (\ref{eq:B+-})
reads
\be
{\cal B}^{\pm} = \pm\frac{1}{\Gamma_c}\int_{0}^{\pm\zeta}du \frac{\partial
\ci}{\partial u} \int_{V_{\pm}}d\vec{r}\,'G^{\pm}(\vec{r},\vec{r}\,')
\frac{\partial c^{\rm in}(\vec{r}\,',t)}{\partial t}.
\ee
We note that $\partial c^{\rm in}(\vec{r}\,',t)/\partial t \equiv
v\partial c^{\rm in}/\partial u' = {\cal O}(\epsilon)$ since the normal
interface velocity $v = \ep v_{1} + \cdots$, $c^{\rm in} =
\ci + \ep\delta c^{\rm in}_{1} +\cdots$ and $G^{\pm} = \ep G_{1}^{\pm} +
\cdots$. Naive power counting seems to imply that ${\cal B}^{\pm} = {\cal
O}(\ep^{2})$, but changing variables $\vec{r}\,'\rightarrow \us',\scs'$
yields
\ben
 {\cal B}^\pm
&=& \pm \frac{\xi^2}{\tau\gc} \int_0^{\pm\bar{\zeta}} d\us \int_{V_\pm}
\frac{d\us'  d\scs'}{1+\us'\ep \Curv(\scs')}
\sum_{n=1} \sum_{m=1} \ep^{n+m-1}
\nonumber \\
&&\times
v_n G^\pm_m(\us,\scs;\us',\scs')
\pqpq{\ci}{\us}\pqpq{}{\us'}(\ci +\ep\delta c^{\rm in}_{1} +\cdots )
\nonumber \\
&=&
\ep {\cal B}^\pm_1+\ep^2 {\cal B}^\pm_2 + \cdots
\een
where
\ben
 {\cal B}^\pm_1 &=& \pm\frac{\xi^2}{\tau\gc} \int_0^{\pm\bar{\zeta}} d\us
\int_{V_\pm} \ d\us' d\scs' v_1
G^\pm_1
\pqpq{\ci}{\us}\pqpq{\ci}{\us'}
\nonumber \\
 {\cal B}^\pm_2 &=& \pm\frac{\xi^2}{\tau\gc} \int_0^{\pm\bar{\zeta}} d\us
\int_{V_\pm} \ d\us' d\scs'
\Bigl[(v_2 G^\pm_1 + v_1 G^\pm_2
\nonumber \\
&-& v_1\Curv \us'G^\pm_1)\pqpq{\ci}{\us}\pqpq{\ci}{\us'} + v_1
G^\pm_1 \pqpq{\ci}{\us}\pqpq{\delta c^{\rm in}_1}{\us'}\Bigr].
\een
For the Green's function introduced above, ${\cal B}_1$ becomes
\ben
{\cal B}_1 &=& {\cal B}_{1}^{-} + {\cal B}_{1}^{+} = \nonumber \\
&=& -\frac {\xi^2 v_{1}(\scs)}{\tau\Gamma_c}
\int_{-\bar{\zeta}}^{+\bar{\zeta}}d\us[\ci(\us)-\co(\us)]^{2}.
\label{eq:BB}
\een

\section{Two-sided mobility}

	In this appendix the sharp interface equations are 
outlined for a mobility that takes on a constant value in 
each phase. There is the question of consistency of such a theory in the
presence of stochastic noises in the underlying Langevin equations which we
do not attempt to answer. We consider the system in the unphysical limit of
zero noise. For convenience, the mobility in the
phase in the region $u<0$ ($u>0$) is denoted $\gc^-$ ($\gc^+$).
 
	When the mobility $\gc$ is different in the two phases, the
equation of motion for the concentration $c$ becomes
\ben
\label{eq:start22}
\dqdt{c} &=& \vec{\nabla} \left(\gc \cdot \vec{\nabla} \fqfq{\Free}{c}\right)
= \vec{\nabla} \left(\gc \cdot \vec{\nabla}
\delta\mu\right) \nline
&=&\gc\nabla^2\delta\mu + (\vec{\nabla}\gc)\cdot\vec{\nabla}\delta\mu.
\een
The procedure outlined in the main text gives
\ben
\label{eq:greqc2}
 \int_{V_\pm} &d\vec{r}\,'&  \frac{G^{\pm}(\vec{r},\vec{r}\,')}{\gc'} \,
\frac {\partial c'}{\partial t}
= \delta\mu   \nline
&-& \int_{S_{\pm}} d\vec S' \cdot  \Bigl(\delta \mu' \vec{\nabla}' G^\pm -
G^\pm \vec{\nabla}' \delta \mu' \Bigr)\nline
&+& \int_{V_\pm}d\vec{r}'
G^{\pm}(\vec{r},\vec{r}\,')
\vec{\nabla}\delta \mu' \cdot \frac{\vec{\nabla}\gc'}{\gc'}.
\een
The last term in Eq.\ (\ref{eq:greqc2}) is ${\cal O}(\epsilon^3)$ and
can be neglected.   It is straightforward to repeat the calculations of
Section (\ref{sec:innerexp}) for the velocity of the interface and for the
chemical potential at the interface.  All results remain the same 
except, the diffusion constant $D_c$ has an obvious dependence 
on $\Gamma_c^{\pm}$ and 
\ben
{\cal E}^2 = &&\Delta c\int_0^\infty 
\frac{du}{\gc^+}\left[\co\left(u\right)
-\ci\left(u\right)\right] \nline
&&+\int_{-\infty}^0\frac{du}{\gc^-}
\left[\ci\left(u\right)-\co\left(u\right)\right]^2 \nline 
&&+\int_0^{\infty}\frac{du}{\gc^+}
\left[\ci\left(u\right)-\co\left(u\right)\right]^2
\een
and the interface position $u=0$ is determined by the solvability condition
\be
\int_0^\infty\frac{du}{\gc^+}\left[\co\left(u\right) - \ci\left(u\right)\right] 
=\int_{-\infty}^0\frac{du}{\gc^-}\left[\ci\left(u\right) - \co\left(u\right)\right] 
\ee


\begin{references}
\bibitem{caution}  This Green's function is 
useful only if all the appropriate terms do not  have a constant 
piece.
\bibitem{twod}  In two dimensions there is a logarithmic correction to
this result.
\bibitem{GOLDSTONE} We integrate over the equations using, for example,
$\partial \psi^i/\partial u$ because this is the variable conjugate to
the linear displacement of the surface, that is, the Goldstone mode.
\bibitem{mathtip} The last equality in Eqs.\ (\ref{eq:cex}) and 
(\ref{eq:pex}) was obtained using the relationships
\ben
\pqpq{}{\us} \left(\pqpq{f(c,\psi)}{c}\right) &=&
\pqpq{^2f(c,\psi)}{c^2}\pqpq{c}{\us} +
\pqpq{^2f(c,\psi)}{c\partial \psi}\pqpq{\psi}{\us}
\nonumber \\
&=& \ff{2}{0} \pqpq{c}{\us} + \ff{1}{1} \pqpq{\psi}{\us}
\nonumber 
\een
and
\ben
\pqpq{}{\us} \left(\pqpq{f(c,\psi)}{\psi}\right)  &=&
\pqpq{^2f(c,\psi)}{\psi^2}\pqpq{\psi}{\us} +
\pqpq{^2f(c,\psi)}{c\partial \psi}\pqpq{c}{\us}
\nonumber \\
&=& \ff{0}{2} \pqpq{\psi}{\us} + \ff{1}{1} \pqpq{c}{\us}.
\nonumber
\een
\bibitem{velocity} In two dimensions the tangential component 
of the time derivative can be written 
$v_{T}\partial/\partial s$, where $v_{T}$ accounts for 
flow of concentration (or $\psi$) along the interface. 
$v_T$ must be ${\cal O}(\epsilon)$ since it is zero 
when $\epsilon=0$.
Furthermore in the inner region $s$ is scaled with $\epsilon$
so $v_{T}\partial/\partial s \sim {\cal O}(\epsilon^2)$.  
When this term operates on $c$ or $\psi$ is becomes 
${\cal O}(\epsilon^3)$, since both $c$ and $\psi$ are independent 
of $s$.  The time derivative in the moving frame 
$\partial/\partial t|_{(u,s)}$ is also at least 
${\cal O}(\epsilon^3)$.  This term accounts for the 
relaxation of fluctuations around the steady state profile 
not due to motion of the interface.  For the concentration
field fluctuations relax as $t \sim ({\rm length})^{2}$ which 
implies $\partial/\partial t|_{(u,s)} \sim {\cal O}(\epsilon^2)$.
As with the tangential component this term becomes  
${\cal O}(\epsilon^3)$ when operating on $c$.  
For $\psi$, fluctuations relax exponentially fast 
and thus this term can be neglected.
\end{references}
\end{document}